\documentclass[12pt]{iopart}
\usepackage[T1]{fontenc}
\usepackage{bm,dcolumn}
\expandafter\let\csname equation*\endcsname\relax
\expandafter\let\csname endequation*\endcsname\relax
\usepackage{tabu}
\usepackage{mathrsfs}
\usepackage{graphicx}
\usepackage{verbatim}
\usepackage[section]{placeins}
\usepackage{color}
\usepackage{soul}
\usepackage{xparse}
\usepackage{physics}
\usepackage{siunitx}
\usepackage{soul}
\usepackage{subcaption}
\usepackage{placeins}
\usepackage{array}
\graphicspath{ {./Figures/} }
\usepackage{multirow}
\usepackage{hyperref}
\hypersetup{
    colorlinks=true,
    linkcolor=magenta,
    urlcolor=blue,
    citecolor=red,
}
\usepackage{enumitem}

\newcommand{\RN}[1]{%
  \textup{\uppercase\expandafter{\romannumeral#1}}%
}

\begin{document}
\title{Floquet engineering of optical lattices with spatial features and periodicity below the diffraction limit}
\author{S. Subhankar$^{1,\dagger}$, P. Bienias$^{1}$, P. Titum$^{1,2,3}$, T-C. Tsui$^{1}$, Y. Wang$^{1}$, A. V. Gorshkov$^{1,2}$, S. L. Rolston$^{1}$, J. V. Porto$^{1}$}
\address{$^{1}$Joint Quantum Institute, National Institute of Standards and Technology 
		and the University of Maryland, College Park, Maryland 20742 USA}
\address{$^{2}$Joint Center for Quantum Information and Computer Science, NIST/University of Maryland, College Park, Maryland 20742 USA}
\address{$^{3}$Johns Hopkins University Applied Physics Laboratory, Laurel, Maryland 20723, USA}
\ead{sarthaks@umd.edu$^\dagger$}
\begin{abstract}

Floquet engineering or coherent time periodic driving of quantum systems has been successfully used to synthesize Hamiltonians with novel properties. In ultracold atomic systems, this has led to experimental realizations of artificial gauge fields, topological band structures, and observation of dynamical localization, to name just a few. Here we present a Floquet-based framework to stroboscopically engineer  Hamiltonians with spatial features and periodicity below the diffraction limit of light used to create them by time-averaging over various configurations of a 1D optical Kronig-Penney (KP) lattice. The KP potential is a lattice of narrow subwavelength barriers spaced by half the optical wavelength ($\lambda/2$) and arises from the nonlinear optical response of the atomic dark state. Stroboscopic control over the strength and position of this lattice requires time-dependent adiabatic manipulation of the dark-state spin composition. We investigate adiabaticity requirements and shape our time-dependent light fields to respect the requirements. We apply this framework to show that a $\lambda/4$-spaced lattice can be synthesized using realistic experimental parameters as an example, discuss mechanisms that limit lifetimes in these lattices, explore candidate systems and their limitations, and treat adiabatic loading into the ground band of these lattices.   

\end{abstract}
\maketitle

\section{Introduction} 

 Time-dependent forcing of quantum systems is ubiquitous in quantum mechanics. Small amplitude driving of a quantum system probes its linear response~\cite{altland_simons_2010} while strong driving allows for Hamiltonian engineering~\cite{Eckardt2015,Eckardt2017,Bukov2015,Holthaus2015,Oka2018}. Optical potentials and in particular optical lattices have proven to be a powerful tool for manipulating  ultracold atomic systems and are used in a wide range of experiments~\cite{lewenstein2017ultracold,Dutta_2015,Gross995}. However, the spatial features and periodicity of these potentials (generally arising from the second order ac-Stark shift)  in the far field are constrained by the diffraction limit to be of order the wavelength of light used to create them. In particular, the Fourier decomposition of the far-field optical potential cannot have components with wavelength less than $\lambda/2$,  and thus the minimum lattice spacing is $\lambda/2$. As the lattice spacing determines many of the energy scales in cold-atom lattice systems, it has been of interest to produce optical lattices with smaller spacings in order to increase relevant energy scales~\cite{Nascimbene2015,Lewenstein2007}. Approaches to making subwavelength spaced optical lattices have been proposed~\cite{Dubetsky2002} and realized~\cite{Ritt2006} based on multiphoton effects, and on adiabatic dressing of different spin dependent lattices~\cite{Yi2008}.

Recently, optical lattices based on the nonlinear optical response of dark states~\cite{acki2016,Jendrzejewski2016} were realized~\cite{Wang2018} with $\lambda/2$ periodicity but strongly subwavelength structure within a unit cell, consisting of a  Kronig-Penny-like (KP) lattice of narrow repulsive barriers of width $\simeq \lambda/50$.  Time averaging a stroboscopically applied lattice potential with high spatial frequency Fourier components can give rise to an average potential with periodicity and spatial features less than $\lambda/2$~\cite{Nascimbene2015}. Since the dark-state KP lattice has high spatial frequency Fourier components, it is a candidate progenitor 
lattice with which to realize such a time-averaged,  subwavelength-featured lattice. Here, we explore the implementation of a time-averaged dark-state KP lattice, taking into account realistic imperfections in the dark-state system.   After careful consideration of the adiabaticity requirements, we show that lattices with $\lambda/4$ period can be realized as an example, and discuss the prospects for  lattices with smaller spacing and features. Ref.~\cite{Mateusz2019} explores related ideas about painting arbitrary  subwavelength optical potentials.

In the time-averaged approach, a time-periodic progenitor potential $W_{\text{0}}(x,t)$ is applied such that the atoms experience the time-averaged potential ${W}_{\text{avg}}(x)$:
\begin{equation}
    {W}_{\text{avg}}(x)=\frac{1}{T}\int^{T/2}_{-T/2} W_{\text{0}}(x,t) dt,
    \label{eq:baseidea}
\end{equation}
where $T = 2 \pi / \omega_T$ is the period of $W_{\text{0}}(x,t)$ and $\omega_T$ is the Floquet frequency. In order to successfully realize ${W}_{\text{avg}}(x)$ while avoiding heating, $\omega_T$ must be much faster than the timescale associated with the motional degree of freedom in the lattice, which is set by the energy gaps between bands in the lattice~\cite{Rahav2003,Nascimbene2015}. This requirement suggests that $\omega_T$ be as large as possible.  
As we discuss below, the particular realization of $W_{\text{0}}(x,t)$ using a dark-state lattice~\cite{Wang2018} has an additional  requirement of spin adiabaticity that limits the maximum allowable 
$\omega_T$.

The dark-state lattice is an artificial scalar gauge potential~\cite{Moody1989,Dalibard2011,Wang2018,Jendrzejewski2016,acki2016} experienced by an atom in the dark-state eigenfunction of a three-level $\Lambda$-system with a spatially dependent spin composition. Dynamically  manipulating the height, barrier width, and position of the lattice requires time-dependent manipulation of the spin composition of the dark-state eigenfunction. This spin manipulation can be seen as a Stimulated Raman Adiabatic Passage (STIRAP) process~\cite{Vitanov2017} and adiabaticity requirements set an upper bound on the window for usable $\omega_T$ within which the atoms are simultaneously motionally diabatic and spin adiabatic. Understanding the practical limits of these constraints requires a detailed consideration of the system dynamics, which we apply to the specific $^{171}$Yb system previously used to demonstrate the dark-state lattice~\cite{Wang2018}.

\section{Time-dependent dark-state potentials} 
\begin{figure*}[]
   \centering
\includegraphics[height=3.3in]{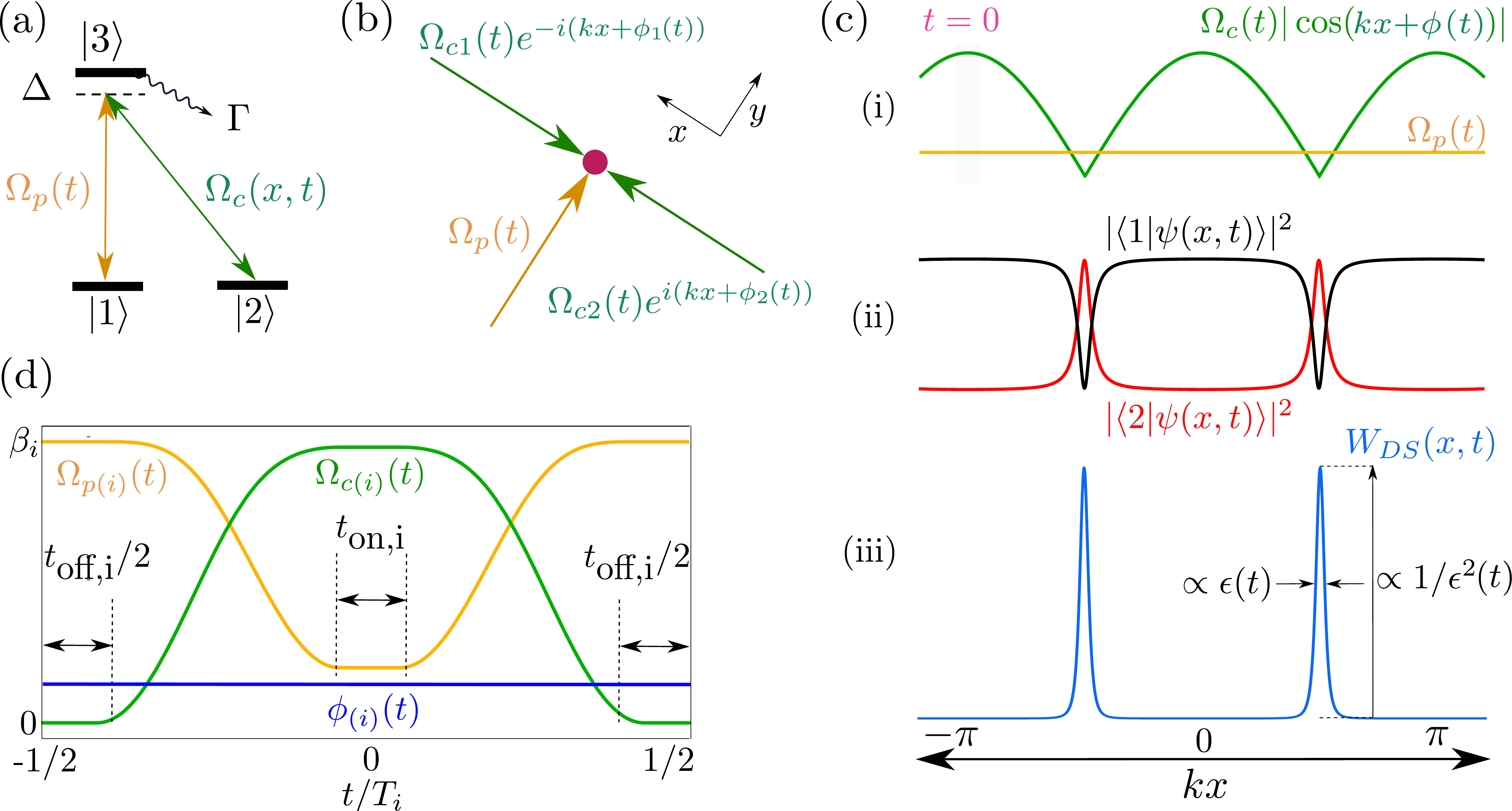}
 \caption{\textbf{(a)} An ideal $\Lambda$-system with inverse lifetime $\Gamma$ and single-photon detuning $\Delta$. One leg of the $\Lambda$-system is coupled by a spatially homogeneous and temporally varying probe light field $\Omega_p(t)$ and the other leg by a spatially inhomogeneous and temporally varying control light field $\Omega_c(x,t)$. \textbf{(b)} The geometry of the light fields with arbitrary control over the envelope, $\Omega_{c1}(t)$, $\Omega_{c2}(t)$, $\Omega_{p}(t)$ and phase, $\phi_{1}(t)$,  $\phi_{2}(t)$ of each light field. \textbf{(c)} (i) The instantaneous (at $t=0$) spatial dependence of the light fields $\Omega_{c}(t)|\cos(kx+\phi(t))|$ and $\Omega_p(t)$, (ii) the probability densities of the spin composition of the dark-state eigenfunction $|\psi(x,t)\rangle$ i.e. $|\langle 1|\psi(x,t)\rangle|^2$ and $|\langle 2|\psi(x,t)\rangle|^2$, and (iii) the instantaneous shape of $W_{DS}(x,t)$. \textbf{(d)} Typical pulse shapes considered here for the control beams $\Omega_{c(i)}(t)=2\Omega_{c1(i)}(t)=2\Omega_{c2(i)}(t)$, probe beam $\Omega_{p(i)}(t)$, and phase $\phi_{i}(t)$ for the $i$th sub-Floquet period where ${-T_i/2\le t\le T_i/2}$ that determines the time-averaged potential $W_{\textrm{avg}}(x)$. }
 \label{timeKP}
\end{figure*}

We consider the creation of time-periodic potentials for the dark-state channel, $ W_{DS}(x,t) $ (which serves as $W_{\text{0}}(x,t)$ in Eq.~\ref{eq:baseidea}),
by coupling the three atomic levels in a $\Lambda$-system with a spatially homogeneous probe light field $\Omega_p(t)$, and a spatially inhomogeneous control light field. The inhomogeneous control light field is composed of two counter propagating fields with equal magnitudes driven simultaneously, $\Omega_c(x,t)=\Omega_c(t)\cos(kx+\phi(t))$ where $k=2\pi/\lambda$, as shown in Fig.~\ref{timeKP}{\color{magenta}a}. Working in the spatially and temporally local dressed state basis of the $\Lambda$-system determined by the coupling fields $\Omega_p(t)$ and $\Omega_c(x,t)$, the Hamiltonian is given by (Eq.~\ref{Heff})
\begin{align}
    \hat{H}_{\text{rot}}(x,t)&=\frac{\hat{p}^2}{2m}+\left(
\begin{array}{ccc}
W_{DS}(x,t) & 0 &
  0 \\
 0 & W_-(x,t) &0 \\
0 &
   0 & W_+(x,t) \\
\end{array}
\right)+\hat{H}_{\text{od}}(x,p,t),
\end{align}
where $W_{DS}(x,t)$ and $W_{\pm}(x,t)$ are the dark-state and bright-state potentials in the three Born-Oppenheimer (BO) channels and $\hat{H}_{\textrm{od}}(x,p,t)$ represents the 
off-diagonal couplings between these channels (See~\ref{Section2}). $W_{DS}(x,t)$ and $W_{\pm}(x,t)$ include the BO potentials as well as the non-adiabatic corrections to these potentials. The dressed state coupling induced by $\hat{H}_{\textrm{od}}(x,p,t)$ is detrimental, since it mixes bare excited state $|3\rangle$ into the dark-state channel through the bright-state channels, inducing photon scattering in the otherwise lossless dark state.

The spin-composition of the dark-state eigenfunction for the $\Lambda$-system in Fig.~\ref{timeKP}{\color{magenta}a} is ${|DS(x,t)\rangle=-\cos\alpha(x,t)|1\rangle+\sin\alpha(x,t)|2\rangle}$ where $\alpha(x,t)=\tan^{-1}[\Omega_p(t)/\Omega_c(x,t)]$. The non-adiabatic correction to the dark-state BO potential that gives rise to $W_{DS}(x,t)$ is determined by the spatial gradient of the spin composition~\cite{acki2016,Jendrzejewski2016} (\ref{gaugepotderivation}) (Fig.~\ref{timeKP}{\color{magenta}c}),
\begin{equation}
W_{DS}(x,t)=\frac{\hbar^2}{2m}\bigg(\frac{\partial }{\partial x}\alpha(x,t)\bigg)^2,
\label{DSgaugePomain}
\end{equation}
which for the light field configuration considered here is a lattice of narrow repulsive barriers with temporally modulated strength and position. We take here a stroboscopic approach, where $W_{DS}(x,t)$ is repeatedly pulsed on and off in magnitude at $N$ different positions for time $T_i$ with the position of $W_{DS}(x,t)$ being shifted in between the lattice pulses (Here $T=\sum T_i$). In addition, $W_{DS}(x,t)$ can be held on or off for $t_{\text{on,i}}$ and $t_{\text{off,i}}$
(Fig.~\ref{timeKP}{\color{magenta}d}). Time averaging over the $N$ different pulsed KP lattice potentials with arbitrary strength and position can produce an  arbitrary time-averaged potential ${W}_{\text{avg}}(x)$~\cite{Mateusz2019}. 

The ability to paint potentials requires real-time control over the position, strength and width of the barriers (Eq.~\ref{DSgaugePomain}). The strength of the barriers can be controlled via the Rabi frequencies $\Omega_p(t)$ and $\Omega_c(t)$ (Figs.~\ref{timeKP}{\color{magenta}b},~\ref{timeKP}{\color{magenta}c}) with the height and width of the barriers being proportional to $1/\epsilon^2(t)$ and $\epsilon(t)$ respectively~\cite{Wang2018,Jendrzejewski2016,acki2016} where $\epsilon(t)=\Omega_p(t)/\Omega_c(t)$ (for $\epsilon(t)\ll 1$). The barriers are located at the nodes/minimums of $\Omega_c(x,t)$ (Fig.~\ref{timeKP}{\color{magenta}c}), and their positions can be controlled by the differential control beam phase $\phi(t)=\phi_1(t)-\phi_2(t)$ (Figs.~\ref{timeKP}{\color{magenta}b},~\ref{timeKP}{\color{magenta}c}). Stitching $N$ different sub-Floquet periods together (while ensuring continuity in the Rabi pulses between the sub-Floquet periods) into one Floquet period allows for versatility in the time-averaged potential ${W}_{\text{avg}}(x)$ that can be generated. Each sub-Floquet period of duration $T_i$ pulses a KP potential at a different position $x_{0i}$ (determined by the phase $\phi_{0i}$) with a strength and width determined by $\epsilon_i$. Fig.~\ref{timeKP}{\color{magenta}d} shows the pulses $\Omega_{p(i)}(t)$, $\Omega_{c(i)}(t)$ and $\phi_{(i)}(t)$ for the $i$th sub-Floquet period
${-T_i/2\le t\le T_i/2}$.

\section{Adiabaticity considerations}

Without explicit time-dependence, $\hat{H}_{\text{od}}(x,p,t)$ has static, off-diagonal terms depending on the spatial gradient of the dark-state spin composition that couple the dark-state channel to the lossy bright-state channels. This loss mechanism was theoretically~\cite{acki2016,Jendrzejewski2016} and experimentally~\cite{Wang2018} shown to limit lifetimes in the KP lattice. Large energy gaps between the BO channels via large Rabi frequencies $\Omega_c$ and $\Omega_p$ generally aid in suppressing this loss~\cite{Wang2018}. When explicit time dependence to the Rabi frequencies is included i.e. $\Omega_c(x,t)$ and $\Omega_p(t)$, the spatially dependent loss mechanism has a trivial time dependence due to the temporally periodic nature of the changing dressed states, and the loss is quantified by averaging over one Floquet period. There is, however, an additional loss mechanism mediated via an explicitly time-dependent term in $\hat{H}_{\text{od}}(x,p,t)$ (see~\ref{Section2}). This term   mediates non-adiabatic couplings between the dark-state channel and the bright-state channels, but can be suppressed by careful pulse shaping.

 Our goal is to design $\Omega_p(t)$ and $\Omega_c(x,t)$ to be simultaneously motionally diabatic and spin adiabatic. In order to design pulses that are spin adiabatic, we consider the three inequalities that quantify the sufficiency requirements for adiabaticity~\cite{Tong2007} defined at single photon resonance ${\Delta=0}$ (see \ref{sufficiencystirap},~\ref{Pulseshaping2}):  
\begin{subequations}
\begin{eqnarray}
 &\left|\frac{\partial}{\partial t}\alpha(x,t)\right|\ll\Omega_{\textrm{rms}}(x,t),\label{energygap}\\ 
 &\int^{\pi/\omega_T}_{-\pi/\omega_T}\left|\frac{\partial}{\partial t}\left(\frac{\partial\alpha(x,t)/\partial t}{\Omega_{\textrm{rms}}(x,t)}\right)\right|dt\ll1,\label{doublederivative} \\
 &\int^{\pi/\omega_T}_{-\pi/\omega_T}\frac{\left|\partial\alpha(x,t)/\partial t\right|^2}{\Omega_{\textrm{rms}}(x,t)}dt\ll1,\label{rabicriterion}
\end{eqnarray}
\label{adiabaticityrequirements}
\end{subequations}
where $\Omega_{\textrm{rms}}(x,t)=\sqrt{|\Omega_{c}(x,t)|^2+|\Omega_{p}(t)|^2}$. Eq.~\ref{energygap}, called the local adiabatic criterion~\cite{Vitanov2017}, states that  to ensure adiabaticity during pulsing, the energy gap between the dark and bright eigenstates (set by $\Omega_{\textrm{rms}}(x,t)$) must be much greater than the off-diagonal couplings between them ($\left|\partial\alpha(x,t)/\partial t\right|$). Eq.~\ref{doublederivative} forces the pulses to be smooth while both Eqs.~\ref{doublederivative} and~\ref{rabicriterion} set bounds on their rise time and fall times. 

To design pulse shapes that satisfy Eqs.~\ref{energygap}-\ref{rabicriterion}, we parameterize the condition Eq.~\ref{energygap} through a parameter $r(t)$:
\begin{equation}
r(t)=\frac{\partial \alpha(x_{h},t)/\partial t}{ \Omega_{\textrm{rms}}(x_{h},t)},
\label{inequalitymain}
\end{equation}
evaluated at $x=x_{h}$, the position 
where the inequality is the hardest to satisfy. The role of $r(t)$ is to quantify the spin adiabaticity during the rising and falling segments of the $\Omega_{c1}(t)$, $\Omega_{c2}(t)$ and $\Omega_p(t)$ pulses (Fig.~\ref{timeKP}{\color{magenta}d}). Specifying $r(t)$ determines the functional form for the Rabi frequencies and the Floquet frequency.
\begin{figure}[t]
\centering
\includegraphics[height=2.4in]{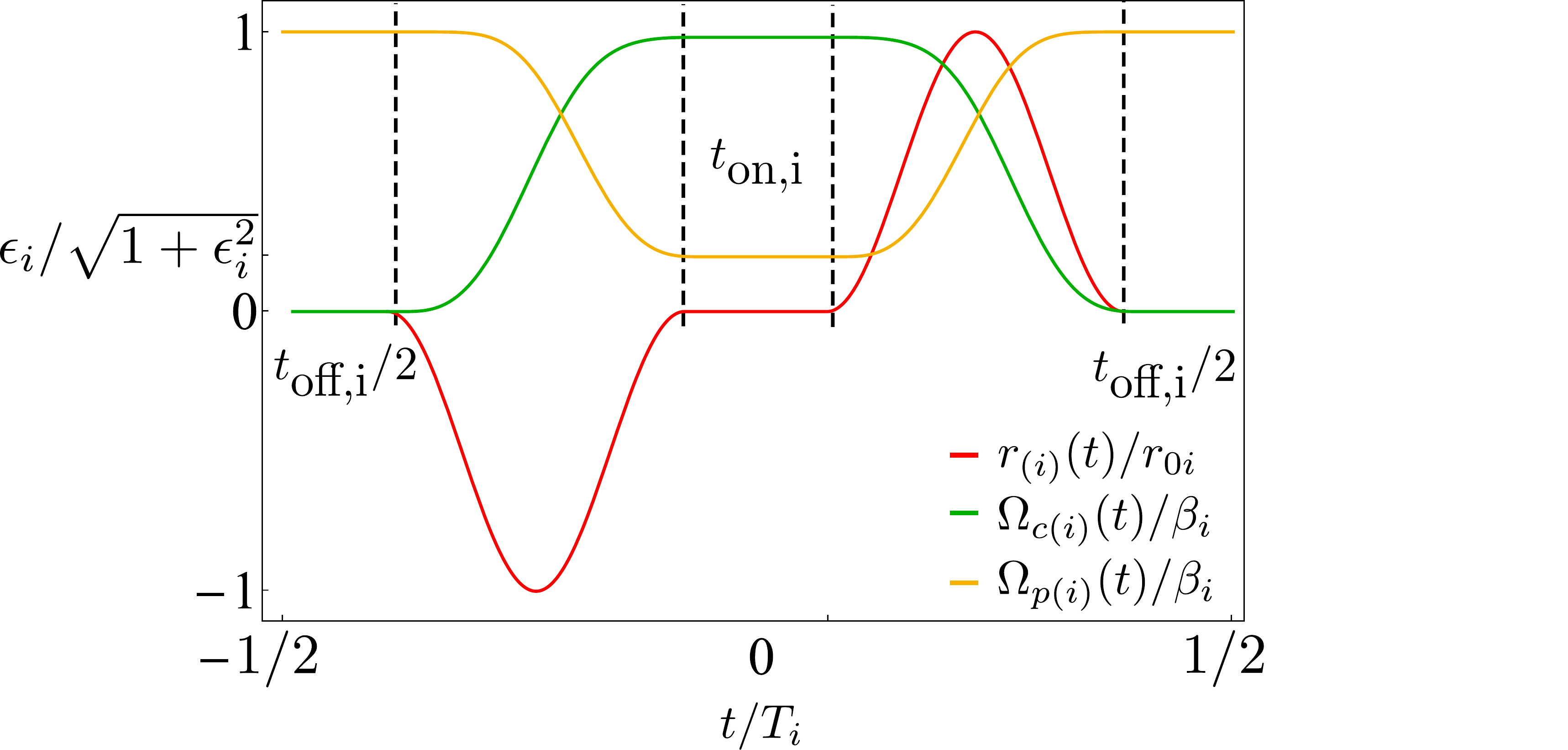}
 \caption{The functional forms for $\Omega_{c(i)}(t)$, $\Omega_{p(i)}(t)$, and $r_{(i)}(t)$ for the $i$th sub-Floquet period ${-T_i/2\le t\le T_i/2}$. Control over the duty cycle of $W_{DS}(x,t)$ is provided by the hold times $t_{\text{off,i}}$ and $t_{\text{on,i}}$.}
\label{gammacombination}
\end{figure}
To satisfy Eq.~\ref{adiabaticityrequirements} during the switching between the on ($t_{\text{on,i}}$) and off times ($t_{\text{off,i}}$), $|r_{(i)}(t)|$ should satisfy the
condition $|r_{(i)}(t)|\ll 1$ (See~\ref{Pulseshaping2}) and  smoothly change from 0.  
We consider a convenient analytic form for $r_{(i)}(t)$ that has a continuous first derivative (Fig.~\ref{gammacombination}):

\begin{equation}
r_{(i)}(t) = \left\{
        \begin{array}{ll}
            0 & \quad -T_i/2\leq t \leq -T_i/2+t_{\text{off,i}}/2 \\
            -r_{0i}\sin^{2}\left(\frac{2\pi (t+t_{\text{on,i}}/2)}{t_{Si}}\right) & \quad -T_i/2+t_{\text{off,i}}/2\leq t \leq -t_{\text{on,i}}/2\\
             0 & \quad -t_{\text{on,i}}/2 \leq t \leq t_{\text{on,i}}/2 \\
            r_{0i}\sin^{2}\left(\frac{2\pi (t-t_{\text{on,i}}/2)}{t_{Si}}\right) & \quad t_{\text{on,i}}/2\leq t \leq T_i/2-t_{\text{off,i}}/2\\
            0 & \quad T_i/2-t_{\text{off,i}}/2\leq t \leq T_i/2
        \end{array}
    \right.
       \label{equalitymain}
\end{equation}
\begin{figure}[]
\centering
\includegraphics[height=3in]{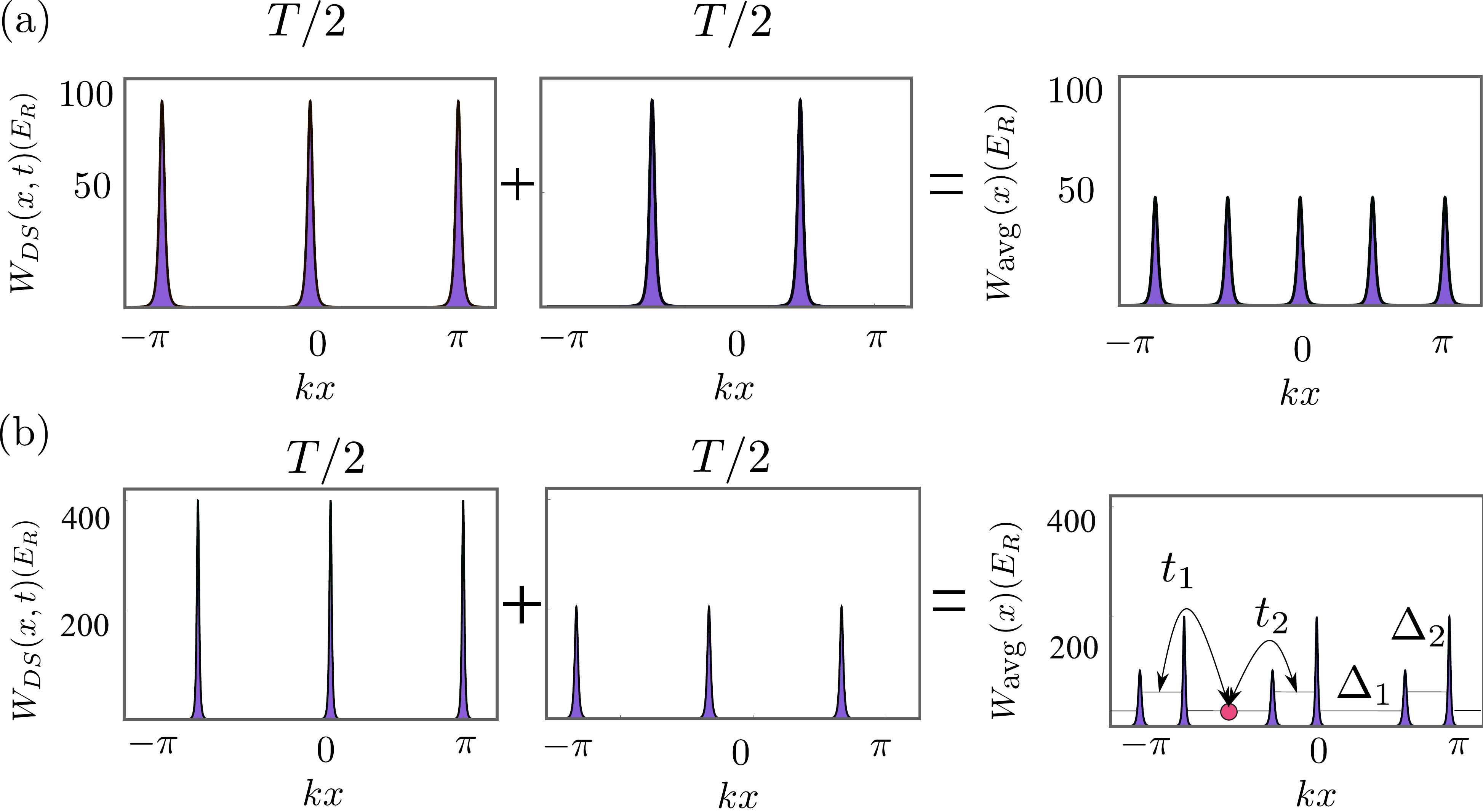}
 \caption{\textbf{(a)} Recipe to stroboscopically realize a $W_{\text{avg}}(x)$ that is a $\lambda/4$-spaced lattice: Pulse $W_{DS}(x,t)$ with $\phi_{(i)}(t)=0$ for $-T/2\le t\le 0$ and $\phi_{(i)}(t)=\pi/2$ for $0\le t\le T/2$. For realistic dark-state lattices, spin adiabaticity requires that the lattice cannot be turned on or off instantaneously. \textbf{(b)} Stroboscopic realization of the Rice-Mele model.}
\label{basicidea}
\end{figure}
where ${T_{i}=t_{Si}+t_{\text{off,i}}+t_{\text{on,i}}}$ and $t_{Si}/2$ is the rise or fall time. 

Generally, it is easiest to be spin adiabatic for large energy separation between the dark and bright-state channels. However at the nodes of $\Omega_c(x,t)$, this energy gap is the smallest with a value of  $\hbar\Omega_p(t)/2$ for $\Delta=0$. Therefore we consider pulse schemes that change the positions of the nodes only when the energy gap at the nodes is large and the spin composition is essentially homogeneous ($\epsilon(t)\gg1$). We consider two ways to achieve the homogeneous condition in between pulses:
\begin{enumerate}[label=(\arabic*)]
    \item $\Omega_p(t)\gg\Omega_c(x,t)$ achieved by turning up $\Omega_p$ while turning off both control beams,
      \item $\Omega_{c1}\gg\Omega_p\gg\Omega_{c2}(t)$ achieved by turning off $\Omega_{c2}(t)$ while $\Omega_p$ and $\Omega_{c1}$ are kept constant.
\end{enumerate}
We note that the pulsing schemes considered here are not unique. Control over $\Omega_{c1}(t)$, $\Omega_{c2}(t)$, $\Omega_{p}(t)$, $\phi_{1}(t)$, and $\phi_{2}(t)$ allows for multiple ways by which arbitrary potentials can be painted, and we refer the reader to Ref.~\cite{Mateusz2019} for other variants. 

For pulse scheme (1), the position where the local adiabatic criterion is the hardest to satisfy,
$x_{h}$, occurs between the nodes. (Pulse scheme (2), for which only one of the two control beams is driven, is treated in the Appendix.) We choose the rms average of the Rabi frequencies to be constant at $x_h$, 
$\Omega_{\text{rms}}=\Omega^2_{p(i)}(t)+\Omega^2_{c(i)}(t)=\beta_i^2=(\Omega^0_{ci})^2+(\Omega^0_{pi})^2=(\Omega^0_{ci})^2(1+\epsilon^2_i)$ where $\beta_i$ is a constant and $\Omega_{c(i)}(t)=2\Omega_{c1(i)}(t)=2\Omega_{c2(i)}(t)$. Solving Eqs.~\ref{inequalitymain} and~\ref{equalitymain} simultaneously, the expressions for  $t_{Si}$, $\Omega_{c(i)}(t)$ are as follows (Fig.~\ref{gammacombination}):
\begin{equation}
\Omega_{c(i)}(t) = \left\{
        \begin{array}{ll}
            0 & \quad -T_i/2\leq t \leq -T_i/2+t_{\text{off,i}}/2 \\
            \beta_i\sin\{\arctan(1/\epsilon_i)\mathcal{G}(t+t_{\text{on,i}}/2)/(2\pi)\} & \quad -T_i/2+t_{\text{off,i}}/2\leq t \leq -t_{\text{on,i}}/2\\
             \Omega^0_{ci} & \quad -t_{\text{on,i}}/2 \leq t \leq t_{\text{on,i}}/2 \\
           \beta_i\sin\{\arctan(1/\epsilon_i)\mathcal{G}(t-t_{\text{on,i}}/2)/(2\pi)\} & \quad t_{\text{on,i}}/2\leq t \leq T_i/2-t_{\text{off,i}}/2\\
            0 & \quad T_i/2-t_{\text{off,i}}/2\leq t \leq T_i/2\\
        \end{array}
    \right.\\
  \label{pulseshape2}
\end{equation}
where $\mathcal{G}(t)=\left|4\pi t/t_{Si}-\sin(4\pi t/t_{Si}) -2\pi|t|/t\right|$ and
\begin{equation}
    t_{Si}= 4\arctan(1/\epsilon_i)/(r_{0i}\beta_i).\label{floquetfrequency2}
\end{equation}     

As a specific example, we explore creation of $\lambda/(2N)$-spaced lattices where $N=2,3,4\ldots$ . These lattices are created by time-averaging $N$ $\lambda/2$-spaced progenitor KP lattice potentials, each shifted in position by $i \lambda/(2N)$ for $(i-1)T/N\le t\le i T/N$ and pulsed for a period of $T_i=T/N$~\cite{Nascimbene2015}, where $i=0,1,...,N-1$~(Fig.~\ref{basicidea}{\color{magenta}a}). More flexibility is possible by pulsing the progenitor lattice with different strengths and relative positions, realizing for example the Rice-Mele model~\cite{Xiao,Rice1982} as shown in Fig.~\ref{basicidea}{\color{magenta}b}.

The goal to create $\lambda /(2N)$-spaced lattices that significantly confines the ground band sets constraints on the lattice parameters. Without requirements of spin adiabaticity, time averaging the KP potential creates $\lambda/(2N)$-spaced lattices with barriers of maximum average height of $(1/N)E_R/\epsilon^2$. Due to the reduction in the size of the unit cell by $N$, the characteristic energy increases to $ N^2E_R$, which is also approximately the energy of the lowest band in a KP lattice.  Hence for the $\lambda/(2N)$-spaced lattice to provide significant confinement,
\begin{equation}
    N^2E_R<\frac{E_R}{N\epsilon^2}\implies\epsilon<\frac{1}{N^{3/2}}.
    \label{bandhost}
\end{equation}
The barrier height for the $\lambda/(2N)$-spaced lattice is 
$
    {{W}_{\text{avg}}(x_{0i})=\int^{T/2}_{-T/2} W_{DS}(x_{0i},t)/T dt,}
$
which can be controlled by choosing $\epsilon$ (limited by requirements on spin adiabaticity) and $t_{\text{off,i}}$ and $t_{\text{on,i}}$. Of course, non-zero values for $t_{\text{off,i}}$ and $t_{\text{on,i}}$ decrease the Floquet frequency ($\omega_T$) as 
\begin{equation}
    \frac{1}{\omega_T}=\frac{N}{2\pi}\left(t_{Si}+t_{\text{off,i}}+t_{\text{on,i}}\right).
    \label{holdchoice}
\end{equation}
Reducing  $\omega_T$ makes it more difficult to be fully  motionally diabatic, so that the operational window between the two constraints rapidly decreases with increasing $N$.

\section{Solving for the Bloch-Floquet bandstructure}
\label{blochfloquetbandstructuremain}
We solve the Bloch-Floquet bandstructure for our Hamiltonian,
\begin{equation}
    \hat{H}(x,t)=\frac{\hat{p}^2}{2m}+\underbrace{\frac{\hbar}{2}\begin{pmatrix}
0 & 0 & \Omega_p(t)  \\
	0 & 0 & \Omega_c(x,t)\\
	\Omega_p(t) & \Omega_c(x,t) & -(2\Delta(t)+i\Gamma)\\
	\end{pmatrix},}_{\hat{\Omega}(x,t)}\\
	\label{Hfullmain}
\end{equation} where $\hat{\Omega}(x+\lambda,t)=\hat{\Omega}(x,t+T)=\hat{\Omega}(x,t)$ with $\Delta(t)=0$ and $\Gamma=48.2\omega_R$ (for the $(6s^2)^1S_0\leftrightarrow(6s6p){}^3P_1$ transition in ${}^{171}$Yb). We substitute the Bloch-Floquet ansatz, $|\psi(x,t)\rangle=e^{iqx-iE_q t/\hbar}|u_{q,E_q}(x,t)\rangle\rangle$~\cite{Holthaus2015,Shirley1965,Hanggi,Eckardt2017,Eckardt2015,Breuer1989,Drese1999} into the time-dependent Schrodinger equation $\hat{H}(x,t)|\psi(x,t)\rangle=i\hbar \frac{\partial}{\partial t}|\psi(x,t)\rangle$ to yield,
\begin{equation}
\hat{K}_q|u_{q,E_q}(x,t)\rangle\rangle=E_q|u_{q,E_q}(x,t)\rangle\rangle
     \label{kamiltonian},
\end{equation}
where $q$ is the quasimomentum, $E_q$ is the quasienergy, $|u_{q,E_q}(x,t)\rangle\rangle$ is the Bloch-Floquet mode, and
$$
    \hat{K}_q=\frac{(\hat{p}+\hbar q)^2}{2m}-i\hbar \frac{\partial}{\partial t}+\hat{\Omega}(x,t),
$$
is the quasienergy operator defined in an extended Hilbert space where time is treated as a coordinate with periodic boundary conditions~\cite{Eckardt2015,Eckardt2017,Holthaus2015}. The extension of the Hilbert space is symbolically represented by the double ket notation of the Bloch-Floquet mode $|u_{q,E_q}(x,t)\rangle\rangle$~\cite{Eckardt2015,Eckardt2017}. We solve the eigenvalue problem in Eq.~\ref{kamiltonian} to calculate the Bloch-Floquet bandstructure (\ref{genhamiltoniansection}). 
\section{Results}
 The loss due to the off-diagonal coupling terms in $\hat{H}_{\text{od}}(x,t)$ that arises from the spatial gradient of the dark-state spin composition increases with smaller $\epsilon$~\cite{acki2016,Wang2018,Jendrzejewski2016}. This suggests that in order to generate potentials that have reasonable lifetimes with realistic values for Rabi frequencies, it is desirable to work at as large an  $\epsilon$ as allowed by Eq.~\ref{bandhost}. 
\begin{figure*}
	\centering
	\includegraphics[height=5.1in]{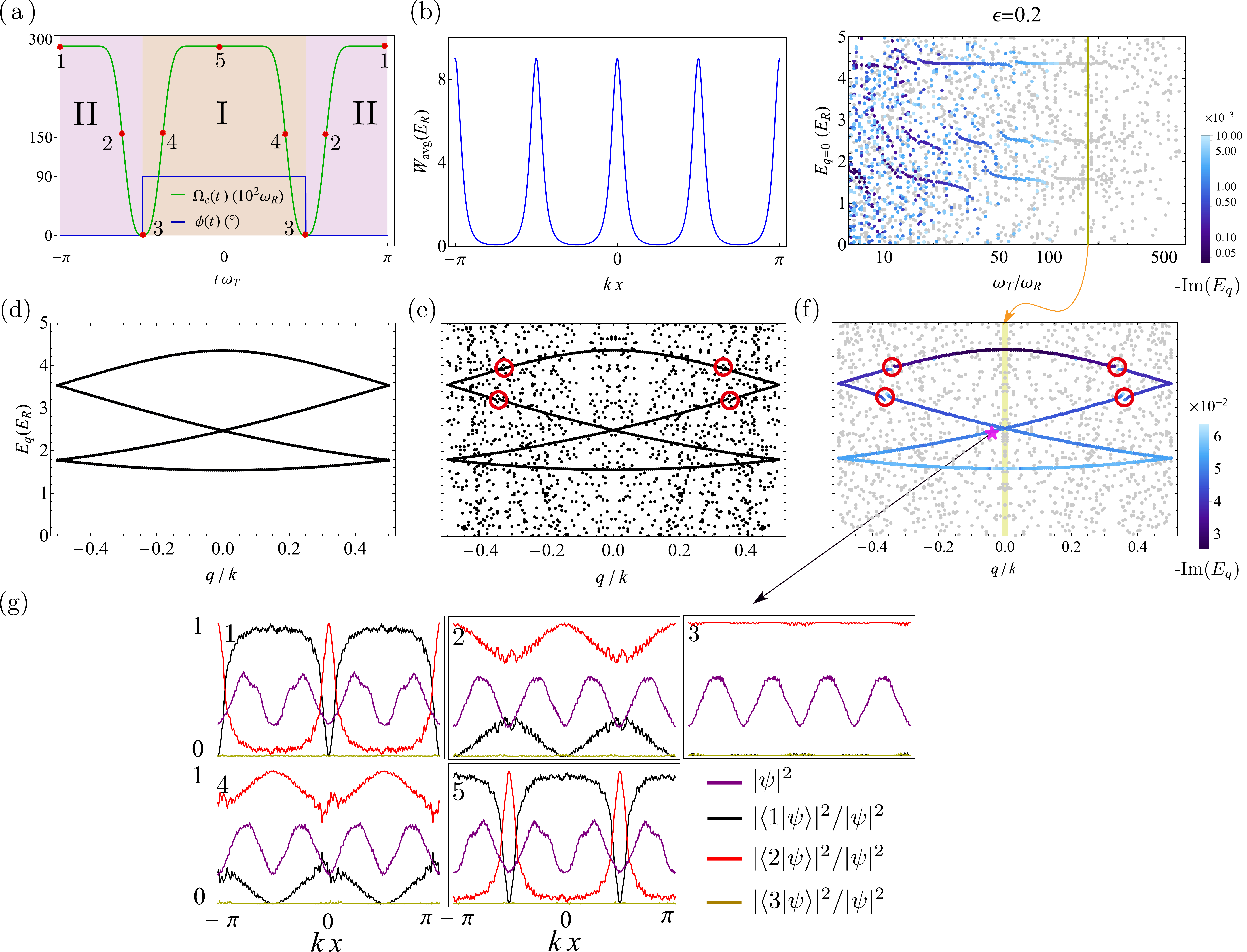}
	\caption{Stroboscopic creation of a $\lambda/4$-spaced lattice potential: \textbf{(a)} One Floquet period constituting the $\Omega_c(t)$ pulse (green trace) and phase pulse $\phi(t)$ (blue trace) for $\epsilon_i=0.2$, $r_{0i}=0.01$, $\Omega^0_{ci}=600\Gamma$, and $t_{Si}=0.36T_{i}$. The numbered red dots enumerate the different timeslices shown in (g). The two sub-Floquet periods are labelled as $\RN{1}$ and $\RN{2}$.  \textbf{(b)} The time-averaged potential $W_{\text{avg}}(x)$. \textbf{(c)} The Floquet spectrum  $E_q$ at $q=0$ as a function of $\omega_T/\omega_R$. The background of seemingly random points represent Floquet states whose quasienergies are only defined up to modulo $\hbar\omega_T$. The loss rate -Im($E_q)$ is given by the colors of the points.
	 \textbf{(d)} The ground band of the time averaged Hamiltonian $\hat{p}^2/2m+W_{\text{avg}}(x)$.
	 \textbf{(e)} The Bloch-Floquet ground band of the dark-state channel Hamiltonian $\hat{H}_{{DS}}(x,t)=\hat{p}^2/2m+W_{DS}(x,t)$ at $\omega_T/\omega_R=150$.
	 \textbf{(f)} The Bloch-Floquet ground band of $\hat{H}(x,t)$ at $\omega_T/\omega_R=150$.   In (d), (e), (f) the bottom edge of the first excited band is at $\sim8E_R$.  \textbf{(g)} The dynamics of the wavefunction $|\psi\rangle=\langle x ,t|u_{q,E_q}(x,t)\rangle\rangle$ of the state indicated by the pink star in subfigure (d) within one Floquet period, sampled at the times indicated in subfigure (a). }%
	\label{calculatebandstructuregamma2}
\end{figure*}
 In Fig.~\ref{calculatebandstructuregamma2}, we investigate the creation of a $\lambda/4$-spaced lattice potential that significantly confines the ground band. 
 With a choice of $\epsilon_i=0.2$, $r_{0i}=0.01$, $\Omega^0_{ci}=600\Gamma$, and $t_{Si}=0.36T_{i}$ (Eq.~\ref{holdchoice}) we use the pulse shape in Eq.~\ref{pulseshape2} to create this potential (Fig.~\ref{calculatebandstructuregamma2}{\color{magenta}a}). The green trace is the pulse profile for $\Omega_c(t)$ (and therefore $\Omega_p(t)=\sqrt{\beta^2-\Omega^2_c(t)}$) while the blue trace is the phase profile $\phi(t)$ during one Floquet period. The numbered red dots  enumerate the different time slices during the pulse. One Floquet period of pulsing involves stitching together two sub-Floquet periods that have a relative phase $\phi_{(i)}(t)$ differing by $\pi/2$. Note that the phase is suddenly switched during an off period when there is no spatial variation to the dark state. The sub-Floquet periods are color coded and labelled as $\RN{1}$ and $\RN{2}$. This pulse yields an effective $\lambda/4$-spaced $W_{\text{avg}}(x)$ potential with $\sim8E_R$ barriers as shown in Fig.~\ref{calculatebandstructuregamma2}{\color{magenta}b} where $W_{\text{avg}}(x)$ is plotted. 
 
 In Fig.~\ref{calculatebandstructuregamma2}{\color{magenta}c} we search for the window of operational $\omega_T$ within which the bands of an effective $\lambda/4$-spaced lattice Hamiltonian are clearly defined by monitoring the Floquet spectrum $E_q$ at $q=0$ as a function of $\omega_T/\omega_R$. As $\omega_T$ is increased, the pulsing becomes more motionally diabatic, but at the cost of increased $r_{0i}$ (Eq.~\ref{floquetfrequency2}). The increased $r_{0i}$ results in stronger admixing of the dark-state channel with the bright-state channels.  The loss rate -Im$(E_q)$ is encoded in the color of the points in Fig.~\ref{calculatebandstructuregamma2}{\color{magenta}c}. The gray dots have loss (-Im$(E_q)$) much larger than the highest value in the color bar. 
 
In Fig.~\ref{calculatebandstructuregamma2}{\color{magenta}d} we plot the bandstructure in the Brillouin zone of $\hat{H}(x,t)$ for the time-averaged Hamiltonian $\hat{p}^2/(2m)+W_{\text{avg}}(x)$  where $W_{\text{avg}}(x)$ is evaluated using Eq.~\ref{eq:baseidea} (Fig.~\ref{calculatebandstructuregamma2}{\color{magenta}b}) for the chosen pulse shape (Fig.~\ref{calculatebandstructuregamma2}{\color{magenta}a}). The folded bandstructure is indicative of a $\lambda/4$-spaced lattice. The Bloch-Floquet bandstructure for the dark-state channel (Fig.~\ref{calculatebandstructuregamma2}{\color{magenta}e}) is obtained by solving the Hamiltonian  $\hat{H}_{{DS}}(x,t)=\hat{p}^2/(2m)+W_{DS}(x,t)$ for $\omega_T=150\omega_R$ (Eq.~\ref{DSgaugePomain}  and see~\ref{trottersection}). The avoided crossings enclosed in the red circles in Fig.~\ref{calculatebandstructuregamma2}{\color{magenta}e} arise from couplings with high-lying dark-state eigenfunctions. The bandstructures shown in  Figs.~\ref{calculatebandstructuregamma2}{\color{magenta}d} and \ref{calculatebandstructuregamma2}{\color{magenta}e} ignore $\hat{H}_{\text{od}}(x,t)$ and therefore exclude loss due to non-adiabatic couplings with the bright states.
In Fig.~\ref{calculatebandstructuregamma2}{\color{magenta}f}, we show the Bloch-Floquet bandstructure of $\hat{H}(x,t)$ (Eq.~\ref{Hfullmain}), which includes the non-adiabatic bright-state couplings.  The avoided crossings exist in the Bloch-Floquet bandstructure at the same place ($q,E_q$) for the same parameters in Fig.~\ref{calculatebandstructuregamma2}{\color{magenta}e}, suggesting that these crossings arise from couplings with high-lying dark-state eigenfunctions. The ground Bloch-Floquet band for $\hat{H}(x,t)$ has the same shape as the static $\lambda/4$-spaced lattice (except near the avoided crossings). The calculated lifetime in the time-averaged potential for the ground band in Fig.~\ref{calculatebandstructuregamma2}{\color{magenta}f} is 1 ms, which can be substantially improved with a lower $\omega_T$ (Fig.~\ref{calculatebandstructuregamma2}{\color{magenta}c}). In general, lifetimes can be increased and the avoided crossings can be removed by operating at larger Rabi frequencies. 

In Fig.~\ref{calculatebandstructuregamma2}{\color{magenta}g} we show the dynamics of the spatial probability densities of the dark-state Bloch-Floquet mode and its spin composition at time slices 1 to 5 (Fig.~\ref{calculatebandstructuregamma2}{\color{magenta}a}) in the $(q,E_q)$ configuration labelled by the pink star in Fig.~\ref{calculatebandstructuregamma2}{\color{magenta}f}. The purple trace is the scaled probability density of the Bloch-Floquet dark-state mode and it has $\lambda/4$ periodicity. It is roughly stationary except for the small wiggles that correspond to micromotion. Meanwhile the spin composition (the black and red traces) of the Bloch-Floquet mode changes dramatically as a function of time during the Floquet period. The population in $|3\rangle$ (yellow trace) remains quite small. 

 \section{Experimental considerations and limitations}
 While working with large Rabi frequencies reduces losses, a significant disadvantage is that the $\Lambda$-system approximation may break down. Perfect $\Lambda$-systems are rare in nature, and $\Omega_p(t)$ and $\Omega_{c}(t)$ can couple off-resonantly to states outside of the $\Lambda$-system. These off-resonant couplings manifest as effective two-photon detunings for the bare ground states of the approximate $\Lambda$-system. Non-zero two-photon detunings are detrimental to STIRAP~\cite{Vitanov2017,Shore2017}, although spatially homogeneous detuning could in principle be compensated with time-dependent laser detuning. Two-photon detunings originating from $\Omega_c(x,t)$, however, are temporally and spatially modulated and may not be completely compensated without significant experimental overhead of adding more spatially dependent compensating laser fields. In addition to added two-photon detuning, the lifetime in the time-averaged lattices is further limited due to the increased excited state admixing with the bare ground states of the $\Lambda$-system. Hence there are trade-offs when increasing the magnitude of the Rabi frequencies: while the dark state evolution is more adiabatic with less bright state admixture, the off-resonant scattering from states outside the $\Lambda$-system also increases. \ref{twophotonsection} presents calculations for a realistic system consisting of ${}^{171}$Yb atoms, which were used to create KP lattices~\cite{Wang2018}. 
 
 A number of techniques can be used to verify the creation of these subwavelength lattices. For example,
nanoresolution spectroscopy~\cite{Subhankar2019} can be used to directly map out the probability density of atoms in the ground band of the $\lambda/(2N)$-spaced lattices. In addition, Bloch oscillations~\cite{Dahan1996} or time-of-flight measurements of the momentum distributions~\cite{Kastberg1995} could be used to measure the $N$ times larger Brillouin zones. Realization of a $\lambda/4$-spaced lattice in $^{171}$Yb is currently being pursued~\cite{Tsui2019}.
 
To create subwavelength spaced lattices with larger $N$, the window of operational $\omega_T$ would be smaller by $N$ (Eq.~\ref{holdchoice}). This requires working with smaller $\epsilon$ as well (Eq.~\ref{bandhost}), making the loss mechanisms that limit lifetime in these lattices more significant. However, higher Rabi frequencies can combat both these limitations, subject to the issues discussed above. 

To adiabatically load into the ground band of the time-averaged $\lambda/(2N)$-spaced lattice potential, the stroboscopic lattice should be turned on slower than the motional time scale set by $\hbar/(N^2E_R)$, while maintaining a large gap to the bright states at all times. For pulse scheme (1) this can be achieved by slowly adjusting the envelope of the pulsed control beam $\tilde{\Omega}_c(t)= f(t) \Omega_c(t)$ while maintaining constant $\Omega_{\text{rms}}$ (See~\ref{adiabaticloading}).

\section{Summary and Outlook}
In this paper, we evaluate the idea of stroboscopically generating  potentials using the repulsive barriers of a dark-state KP potential. We analyzed the competing requirements of maintaining dark state spin adiabaticity and simultaneous motional diabaticity during pulsing of the KP potentials in the presence of realistic imperfections. We showed that it is possible to create such potentials in a experimental system of $^{171}$Yb atoms by calculating the Floquet spectrum of atoms in a stroboscopically  generated $\lambda/4$-spaced lattice. This approach is applicable to any three-level system, although it needs to be well isolated from coupling to other levels, which will shorten lifetimes. While we have treated 1D systems here, this method can be readily generalized to 2D. Using progenitor lattices of subwavelength attractive trapping potentials~\cite{Bienias2018} in conjunction with barriers provides flexibility in tailoring the potential landscape.
\section{Acknowledgements}
	We would like to thank Jean Dalibard for stimulating discussions. S.S., Y.W., T-C.T., S.L.R., and J.V.P. acknowledge support from NSF PFC at JQI (Grant No. PHY1430094) and ONR (Grant No. N000141712411). P.B. and A.V.G. acknowledge funding by AFOSR, ARL CDQI, NSF PFC at JQI, DoE ASCR Quantum Testbed Pathfinder program, ARO MURI, NSF PFCQC program, and the DoE BES Materials and Chemical Sciences Research for Quantum Information Science program. PT acknowledges support from the NRC postdoctoral fellowship.
	
\textit{Note}-We note that related and complementary work is being pursued by {\L}acki et. al.~\cite{Mateusz2019}.
\appendix
\section{Sufficiency conditions for adiabaticity}
\label{Section2}
We start with the time-dependent Hamiltonian,
\begin{equation}
    \hat{H}(x,t)=\frac{\hat{p}^2}{2m}+\frac{\hbar}{2}\begin{pmatrix}
0 & 0 & \Omega_p(t)  \\
	0 & 0 & \Omega^*_c(x,t)\\
	\Omega_p(t) & \Omega_c(x,t) & -(2\Delta(t)+i\Gamma)\\
	\end{pmatrix}\\.
	\label{Hfull}
\end{equation}
$\hat{H}(x,t)$ is non-Hermitian due to the $i\Gamma/2$ term and requires a biorthogonal set of eigenvectors to diagonalize it~\cite{Brody2014}. Due to the non-Hermitian nature of $\hat{H}(x,t)$ the eigenvectors are not guaranteed to be orthogonal to each other, but still form a linearly independent set that spans the Hilbert space~\cite{Brody2014}. To derive the artificial gauge potentials and for quantifying the sufficiency conditions for adiabaticity, we transform $\hat{H}(x,t)$ using a rotation transform $\hat{R}(x,t)$ composed of the right eigenvectors~\cite{Brody2014} of the spin-light field coupling part of $\hat{H}(x,t)$. The expression for $\hat{R}(x,t)$ 
is:
\begin{equation}
\hat{R}(x,t)=\begin{pmatrix}
-\cos\alpha e^{i\theta} & 	\sin\alpha \frac{1}{\sqrt{1+l^2}} & \sin\alpha \frac{1}{\sqrt{1+u^2}}  \\
	\sin\alpha & \cos\alpha \frac{e^{-i\theta}}{\sqrt{1+l^2}}  & \cos\alpha \frac{e^{-i\theta}}{\sqrt{1+u^2}}\\
		0 & \frac{l}{\sqrt{1+l^2}} & \frac{u}{\sqrt{1+u^2}}\\
	\end{pmatrix},\\
	\label{rotationtransform}
\end{equation} where
\begin{align}
    \alpha&=\tan^{-1}\bigg|\frac{\Omega_p(t)}{\Omega_c(x,t)}\bigg|,\quad \zeta=\sqrt{|\Omega_c(x,t)|^2+\Omega^2_p(t)},\nonumber\\\theta&=\text{Arg } \Omega_c(x,t),\quad l=\frac{2E_-(x,t)}{\zeta},\quad u=\frac{2E_+(x,t)}{\zeta}.
\end{align}

For pulse scheme (1) where both control beams are changed simultaneously with equal magnitude: $\Omega_c(x,t)=\Omega_c(t)\sin(kx+\phi(t))$ resulting in $\theta(x,t)=0$. For pulse scheme (2) where only one control beam is pulsed: $\Omega_c(x,t)=\Omega_{c2}(t)e^{i(\phi_2(t)+kx)}/2i -\Omega^0_ce^{-ikx}/2i$ resulting {$\theta(x,t)\ne$ constant}. 

The transformation ${\hat{H}_{\text{rot}}(x,t)=\hat{R}^{-1}\hat{H}\hat{R}-i\hbar \hat{R}^{-1}\partial\hat{R}/\partial t}$ rotates $\hat{H}(x,t)$ into the dressed-atom picture of the $\Lambda$-system. The effective Hamiltonian after the transformation is~\cite{Fleischhauer2005,Moody1989,Vitanov2017,Shore2017,Cheneau2008,NGoldmanGJuzeliunas2014} 
\begin{equation}
    \hat{H}_{\text{rot}}(x,t)=\frac{(\hat{p}-\hat{A})^2}{2m}-\hat{B}+\underbrace{\hbar\begin{pmatrix}
0 & 0 & 0  \\
	0 & E_-(x,t) & 0\\
	0 & 0 & E_+(x,t)\\
	\end{pmatrix}}_{\hat{E}_{BO}(x,t)}\\,
	\label{rotcompact}
\end{equation}
where $
    {\hat{A}=i\hbar \hat{R}^{-1}\nabla \hat{R}}
$, $\hat{B}=i\hbar \hat{R}^{-1}\partial\hat{R}/\partial t$,
$\Delta_{\Gamma}(t)=\Delta(t)+i\Gamma/2$, and $E_{\pm}=\big(-\Delta_{\Gamma}(t)\pm\sqrt{\Delta^2_{\Gamma}(t)+\Omega^2_p(t)+|\Omega_c(x,t)|^2}\big)/2$ are the energies of the upper and lower bright states.

We rearrange the terms in Eq.~\ref{rotcompact} to separate the motion of atoms in the three BO channels (dark state, upper-bright state, and lower-bright state)~\cite{Moody1989,acki2016,Jendrzejewski2016} from the off-diagonal couplings ($\hat{H}_\text{od}(x,p,t)$). For pulse scheme (1) (${\theta(x,t)=0}$) this gives  
\begin{align}
    \hat{H}_{\text{rot}}(x,t)=&\overbrace{\frac{\hat{p}^2}{2m}
	+\frac{\hbar^2}{2m}\left(
\begin{array}{ccc}
(\alpha ^{\prime})^2 & 0 &
  0 \\
 0 & \frac{l^{\prime}
   u^{\prime}}{(l-u)^2}-\frac{ u }{(l-u)}(\alpha ^{\prime})^2 &0 \\
0 &
   0 &
   \frac{l^{\prime} u^{\prime}}{(l-u)^2}+\frac{l }{(l-u)}(\alpha ^{\prime})^2 \\
\end{array}
\right)+\hat{E}_{BO}(x,t)}^{\text{Born-Oppenheimer channels}}\nonumber\\&\underbrace{-\hat{B}(x,t)-\frac{\hat{p}.\hat{A}(x,t)}{2m}-\frac{\hat{A}(x,t).\hat{p}}{2m}+\hat{N}(x,t)}_{\hat{H}_\text{od}(x,p,t)},
\label{Heff}
\end{align}
where ${f}^{\prime}=\partial f/\partial x$ and  $\dot{f}=\partial f /\partial t$. The term $\hat{B}(x,t)$ arises from the explicit time dependence of the $\hat{H}(x,t)$. Careful pulse shaping can help suppress the terms in $\hat{B}(x,t)$ that couple the dark-state channel with the bright-state channels. The coupling terms in $\hat{H}_\text{od}(x,p,t)$ depend only on the ratio of the Rabi frequencies ($\alpha^{\prime}, \dot{\alpha}$) and not on their absolute magnitudes, while the energy separation between the channels ($\hat{E}_{BO}(x,t)$) depend on absolute magnitudes of the Rabi frequencies. Thus at higher Rabi frequencies the BO channels become increasingly decoupled.
In addition, $\Delta\ll\Omega^0_p,\Omega^0_{c}$ ensures that the bright-state channels are well separated from the dark-state channel.  
\subsection{Floquet scalar gauge potentials}
\label{gaugepotderivation}
Here we derive the expression for the Floquet scalar gauge potential for the dark-state channel. The expression for $\hat{A}$ 
when both control beams are driven simultaneously as in pulse scheme (1) is 
\begin{equation}
\hat{A}= i\hbar \left(
\begin{array}{ccc}
 0 &- \frac{ 1}{\sqrt{l^2+1}}\alpha ^{\prime} & -\frac{ 1}{\sqrt{u^2+1}}\alpha ^{\prime} \\
 -\frac{\sqrt{l^2+1} u }{(l-u)}\alpha ^{\prime} & 0 &
   -\frac{u^{\prime}\sqrt{l^2+1} }{(u-l) \sqrt{u^2+1}} \\
 -\frac{l \sqrt{u^2+1}  }{
   (u-l)}\alpha ^{\prime} & -\frac{l^{\prime}\sqrt{u^2+1} }{\sqrt{l^2+1} (l-u)} & 0 \\
\end{array}
\right).
\end{equation} 
The expression for the scalar gauge potentials for the BO channels is~\cite{Cheneau2008,Dalibard2011}:

\begin{align}
    \frac{\hat{A}^2-\text{Diag}(\hat{A})^2}{2m}&=
    \frac{\hbar^2}{2m}\left(
\begin{array}{ccc}
(\alpha ^{\prime})^2 & 0 &
  0 \\
 0 & \frac{l^{\prime}
   u^{\prime}}{(l-u)^2}-\frac{ u }{(l-u)}(\alpha ^{\prime})^2 &0 \\
0 &
   0 &
   \frac{l^{\prime} u^{\prime}}{(l-u)^2}+\frac{l }{(l-u)}(\alpha ^{\prime})^2 \\
\end{array}
\right)\nonumber\\&+
\overbrace{\frac{\hbar^2}{2m}\left(
\begin{array}{ccc}0 & -\frac{l l^{\prime} \alpha ^{\prime}}{\left(l^2+1\right)^{3/2}} &
   -\frac{u u^{\prime} \alpha ^{\prime}}{\left(u^2+1\right)^{3/2}} \\
 -\frac{l \sqrt{l^2+1} u^{\prime} \alpha ^{\prime}}{(l-u)^2} & 0 & \frac{\sqrt{l^2+1}
   u (\alpha ^{\prime})^2}{\sqrt{u^2+1} (u-l)} \\
 -\frac{u \sqrt{u^2+1} l^{\prime}\alpha ^{\prime}}{ (l-u)^2} &
   \frac{l \sqrt{u^2+1} (\alpha ^{\prime})^2}{\sqrt{l^2+1} (l-u)} &
   0 \\
\end{array}
\right)}^{\hat{N}(x,t)},
\end{align}
where the first matrix contains the scalar gauge potentials for each of the BO channels. The scalar gauge potential for the dark-state channel is,
\begin{equation}
W_{DS}(x,t)=\frac{\hbar^2}{2m}\bigg(\frac{\partial }{\partial x}\alpha(x,t)\bigg)^2.
\end{equation}
The expressions for $\hat{A}(x,t)$ and $W_{DS}(x,t)$ for pulse scheme (2) ($\theta(x,t)\ne$constant), are
\begin{equation}
\hat{A}= i\hbar \left(
\begin{array}{ccc}
 i \theta^{\prime}\cos ^2\alpha   &- \frac{e^{-i\theta}(i \sin (2 \alpha ) \theta^{\prime}+2 \alpha ^{\prime})}{2\sqrt{l^2+1}} & -\frac{e^{-i\theta}(i \sin (2 \alpha ) \theta^{\prime}+2 \alpha ^{\prime})}{2\sqrt{u^2+1}} \\
 \frac{e^{i\theta}\sqrt{l^2+1} u \left(i \sin (2 \alpha ) \theta^{\prime}-2 \alpha ^{\prime}\right)}{2(l-u)} & \frac{i  u \theta^{\prime} \cos
   ^2\alpha}{ (l-u)} &
   -\frac{\sqrt{l^2+1} \left(i u \theta^{\prime} \cos^2\alpha+u^{\prime}\right)}{(u-l) \sqrt{u^2+1}} \\
 \frac{e^{i\theta}l \sqrt{u^2+1} \left(i \sin (2 \alpha ) \theta^{\prime}-2 \alpha ^{\prime}\right)}{2
   (u-l)} & -\frac{\sqrt{u^2+1} \left(i l \theta^{\prime} \cos^2\alpha+l^{\prime}\right)}{\sqrt{l^2+1} (l-u)} & -\frac{i l 
  \theta^{\prime} \cos ^2\alpha}{(l-u)
 } \\
\end{array}
\right),
\label{genA}
\end{equation} 
and
\begin{align}
   W_{DS}(x,t)&= \frac{\hbar^2}{2m}\bigg[\frac{1}{4}\bigg(\sin(2\alpha(x,t))\frac{\partial }{\partial x}\theta(x,t)\bigg)^2+\bigg(\frac{\partial }{\partial x}\alpha(x,t)\bigg)^2\bigg].
\end{align}

\subsection{Formulating the sufficiency conditions for adiabaticity}
\label{sufficiencystirap}
The general expression for $\hat{B}$ is analogous to $\hat{A}$ except that derivatives are with respect to $x$ in $\hat{A}$ and with respect to $t$ in $\hat{B}$, which for pulse scheme (1) is: 
\begin{equation}
\hat{B}=i\hbar\left(
\begin{array}{ccc}
0 & -\frac{1 }{\sqrt{l^2+1}}\dot{\alpha} & -\frac{1 }{\sqrt{u^2+1}}\dot{\alpha} \\
 -\frac{ \sqrt{l^2+1} u }{ (l-u)}\dot{\alpha} & 0 &
   -\frac{\sqrt{l^2+1} }{(u-l) \sqrt{u^2+1}}\dot{u} \\
 -\frac{ l \sqrt{u^2+1} }{ (u-l)}\dot{\alpha} & -\frac{\sqrt{u^2+1} }{\sqrt{l^2+1} (l-u)}\dot{l} & 0 \\
\end{array}
\right).
\label{Bmatrix}
\end{equation}
The local adiabatic criterion (Eq.~\ref{energygap}) for the instantaneous dark state requires that the minimum of the spatially and temporally varying energy gap between the dark and bright states must be much larger than the largest off-diagonal couplings between them, which we quantify as 
 \begin{equation}
     \text{min}(|E_-(x_{h},t)|,|E_+(x_{h},t)|)\gg\bigg|\frac{\partial}{\partial t}\alpha(x_{h},t)\bigg|.
 \end{equation}
For $\Delta=0$, where fastest STIRAP pulses are guaranteed~\cite{Dykhne1962,Vitanov2017,Shore2017,Vasilev2009}
\begin{equation}
\Omega_{\textrm{rms}}(x_{h},t)\gg\bigg|\frac{\partial}{\partial t}\alpha(x_{h},t)\bigg|, 
\label{localAdia}
\end{equation}
 where $\Omega_{\textrm{rms}}(x,t)=\sqrt{|\Omega_{c}(x,t)|^2+\Omega^2_{p}(t)}$. 
 
The expression for pulse scheme (2) (${\theta(x,t)\ne\text{constant}}$)
is:
\begin{equation}
\hat{B}=i\hbar\left(
\begin{array}{ccc}
 i\dot{\theta} \cos ^2\alpha   & -\frac{e^{-i \theta} \left(i \sin (2 \alpha )
   \dot{\theta}+2\dot{\alpha}\right)}{2\sqrt{l^2+1}} & -\frac{e^{-i \theta} \left(i  \sin (2 \alpha ) \dot{\theta}+2\dot{\alpha}\right)}{2\sqrt{u^2+1}} \\
 \frac{e^{i \theta} \sqrt{l^2+1} u \left(i \sin (2 \alpha ) \dot{\theta}-2 \dot{\alpha}\right)}{2 (l-u)} & \frac{i  u \dot{\theta} \cos^2\alpha}{ (l-u)} &
   -\frac{\sqrt{l^2+1} \left(i u \dot{\theta} \cos^2\alpha+\dot{u}\right)}{(u-l) \sqrt{u^2+1}} \\
 \frac{e^{i \theta} l \sqrt{u^2+1} \left(i \sin (2 \alpha ) \dot{\theta}-2 \dot{\alpha}\right)}{2 (u-l)} & -\frac{\sqrt{u^2+1} \left(i l \dot{\theta} \cos^2\alpha +\dot{l}\right)}{\sqrt{l^2+1} (l-u)} & -\frac{i l
    \dot{\theta} \cos ^2\alpha }{(l-u) } \\
\end{array}
\right),
\end{equation}
which in addition to $\dot{\alpha}$ and $\alpha$ depends on $\dot{\theta}$ and $\theta.$

\section{Pulse Shaping}
\label{Pulseshaping}
\subsection{Pulse scheme (2)}
In the main text, we consider pulse scheme (1) with the constraint ${\Omega^2_{p(i)}(t)+\Omega^2_{c(i)}(t)=\beta_i^2}$. For pulse scheme (2), in which only one control beam is pulsed, $\Omega_{p(i)}(t)=\Omega^0_{pi}$ and $\Omega_{c1(i)}(t)=\Omega^0_{ci}/2$, the pulse shape $\Omega_{c2(i)}(t)$ and $t_{Si}$ are determined by $r_{(i)}(t)$ (Eq.~\ref{equalitymain}) and $\epsilon_i$ as follows: 

\begin{equation}
\Omega_{c2(i)}(t) = \left\{
        \begin{array}{ll}
           0 & \quad -T_i/2\leq t \leq -T_i/2+t_{\text{off,i}}/2 \\
           \frac{\Omega^0_{ci}}{2}-\frac{\Omega^0_{pi}\mathcal{G}(t+t_{\text{on,i}}/2)}{\sqrt{16\pi^2\epsilon_i^2+4\pi^2-\mathcal{G}^2(t+t_{\text{on,i}}/2)}} & \quad -T_i/2+t_{\text{off,i}}/2\leq t \leq -t_{\text{on,i}}/2\\
              \frac{\Omega^0_{ci}}{2} & \quad -t_{\text{on,i}}/2 \leq t \leq t_{\text{on,i}}/2 \\
          \frac{\Omega^0_{ci}}{2}-\frac{\Omega^0_{pi}\mathcal{G}(t-t_{\text{on,i}}/2)}{\sqrt{16\pi^2\epsilon_i^2+4\pi^2-\mathcal{G}^2(t-t_{\text{on,i}}/2)}} & \quad t_{\text{on,i}}/2\leq t \leq T_i/2-t_{\text{off,i}}/2\\
           0 & \quad T_i/2-t_{\text{off,i}}/2\leq t \leq T_i/2\\
        \end{array}
    \right.\\
    \label{onebeamdrive}
\end{equation}
where $\mathcal{G}(t)=\left|4\pi t/t_{Si}-\sin(4\pi t/t_{Si}) \right|$ and
\begin{equation}
   t_{Si}=\frac{4 }{r_{0i}\Omega^0_{pi}\sqrt{4\epsilon_{i}^2+1}}.
\end{equation}
For this scheme, $x_h$ is at the nodes of $\Omega_c(x,t)$ since the energy gap between the dark-state and bright-state channels is the smallest at the nodes and the spin at the node must completely flip from $|2\rangle$ to $|1\rangle$ (Fig.~\ref{timeKP}{\color{magenta}a}) at the end of the pulse. 
\subsection{Verifying the spin-adiabaticity 
\label{Pulseshaping2}
requirements and choice for $r_{0i}$}
We use the off-diagonal coupling terms in Eq.~\ref{Bmatrix} and set $\Delta=0$ to recast the sufficiency conditions in Ref.~\cite{Tong2007} (the inequalities Eq.~\ref{energygap}-\ref{rabicriterion}). The first condition Eq.~\ref{energygap} implies
\begin{align}
&\left|\frac{\partial}{\partial t}\alpha(x_{h},t)\right|\ll\Omega_{\textrm{rms}}(x_{h},t),
\\\implies&r_{0i}\ll 1
\end{align}
 where we have used Eqs.~\ref{inequalitymain} and \ref{equalitymain}. For $r_{0i}=0.01$, this inequality is well satisfied. 
The stronger version~\cite{Tong2007} of the second inequality Eq.~\ref{doublederivative} states:
\begin{align}
&\left|\frac{\partial}{\partial t}\left(\frac{\partial\alpha(x_{h},t)/\partial t}{\Omega_{\textrm{rms}}(x_{h},t)}\right)\right|_{\textrm{max}}t_{Si}\ll1\label{doublederivativestrong}\\\implies& r_{0i}\ll \frac{1}{2\pi}\simeq0.16.
\end{align}
For $r_{0i}=0.01$, this inequality is also well satisfied. 
We note that Eq.~\ref{doublederivativestrong} also enforces that $r(t)$ must be differentiable. 
The stronger version of the third inequality Eq.~\ref{rabicriterion} is~\cite{Tong2007}
\begin{align}
&\left|\Omega_{\textrm{rms}}(x_{h},t)r^2(t)\right|_{\textrm{max}}t_{Si} \ll1\\\implies&
t_{Si}\beta_i r^2_{0i}\ll 1\implies4\arctan(1/\epsilon_{i})\ll1/r_{0i}\textrm{ for pulse scheme (1)}
\end{align}
where we have substituted Eq.~\ref{floquetfrequency2}. Again this inequality is well satisfied for $r_{0i}=0.01$. 
\section{Bloch-Floquet bandstructure}
\subsection{Bandstructure of $\hat{H}(x,t)$}
\label{genhamiltoniansection}

We evaluate the matrix elements of the quasienergy operator $\hat{K}_q$ derived in Sec.~\ref{blochfloquetbandstructuremain}. $\hat{K}_q$ is expressed in dimensionless units $\tilde{x}$ and $\tilde{t}$ where $\tilde{x}=(2\pi/\lambda) x= k x$, $\tilde{t}=(2\pi/T) t=\omega_T t$, $E_R=\hbar^2k^2/(2m)=\hbar\omega_R$, and the tildes over $x$ and $t$ are dropped for convenience, as follows:
\begin{align}
      &\bigg[(-i\partial_{x}+ q)^2+\hat{\Omega}(x,t)-i\omega_T\frac{\partial}{\partial t}\bigg]|u_{q,E_q}(x,t)\rangle\rangle=E_q|u_{q,E_q}(x,t)\rangle\rangle\\
    \implies& \hat{K}_q|u_{q,E_q}(x,t)\rangle\rangle=E_q|u_{q,E_q}(x,t)\rangle\rangle
     \label{dimensionless}.
\end{align}

We expand the Hilbert space of the Bloch-Floquet modes in a plane wave basis: ${| u_{q,E_q}(x,t)\rangle\rangle=\sum_{lmj}c_{lmj}|lmj\rangle}$ where ${\langle xt  |lmj\rangle=e^{ilx}e^{imt}|j\rangle}$. Here $l\in[-L,L]$, $m\in[-M,M]$, and $j\in[1,2,3]$ represents the three spins. The Hilbert space spanned by the basis set is composed of plane waves with the property $\sum_{lmj}|lmj \rangle\langle lmj|=I^{(2L+1)}\otimes I^{(2M+1)}\otimes I^{3}$. We solve Eq.~\ref{dimensionless} by diagonalizing $\hat{K}_q$. The matrix elements of the spin-independent components of $\hat{K}_q$ $(\langle l^{\prime}m^{\prime}j^{\prime}|\hat{K}_q|lmj \rangle)$ are:

\begin{align}
    \bigg\langle l^{\prime}m^{\prime}j^{\prime} \bigg|-i\frac{\partial}{\partial t} \bigg|lmj \bigg\rangle=\delta_{ll^{\prime}}\otimes(m\delta_{mm^{\prime}})\otimes\delta_{jj^{\prime}},
    \label{timederiv}
\end{align}
\begin{align}
    \bigg\langle l^{\prime}m^{\prime}j^{\prime} \bigg|\left(-i\partial_{x}+ q\right)^2\bigg|lmj \bigg\rangle=\left[\left(l+q\right)^2\delta_{ll^{\prime}}\right]\otimes\delta_{mm^{\prime}}\otimes\delta_{jj^{\prime}}.
    \label{momentumderiv}
\end{align}
The spin-dependent component of $\hat{K}_q$ is $\hat{\Omega}(x,t)$:
\begin{align}
    &\hat{\Omega}(x,t)=\frac{1}{2}\begin{pmatrix}
0 & 0 & \Omega_p(t)  \\
	0 & 0 & \Omega^*_c(x,t)\\
	\Omega_p(t) & \Omega_c(x,t) & -(2\Delta(t)+i\Gamma)\\
	\end{pmatrix}\\                                                                                           &=\frac{1}{2}\bigg (\underbrace{\Omega_p(t)(|1\rangle\langle3|+|3\rangle\langle1|)}_\text{A}+\underbrace{-(2\Delta(t)+i\Gamma)|3\rangle\langle3|}_\text{B}+\underbrace{\Omega^*_c(x,t)|2\rangle\langle3|+\Omega_c(x,t)|3\rangle\langle2|}_\text{C}\bigg),
\end{align}
where the matrix elements of A and B are,
\begin{align}
&\langle l^{\prime}m^{\prime}j^{\prime} |\text{A}|lmj \rangle=\delta_{ll^{\prime}}\otimes\langle m^{\prime}|\Omega_p(t)|m\rangle\otimes(\delta_{j^{\prime}1}\delta_{j3}+\delta_{j^{\prime}3}\delta_{j1}),\\&
\langle l^{\prime}m^{\prime}j^{\prime} |\text{B}|lmj \rangle=-i\Gamma\delta_{ll^{\prime}}\otimes\delta_{mm^{\prime}}\otimes\delta_{j^{\prime}3}\delta_{j3}-2\delta_{ll^{\prime}}\otimes\langle m^{\prime}|\Delta(t)|m\rangle\otimes\delta_{j^{\prime}3}\delta_{j3}.
\end{align}
Depending on the pulse scheme, $\langle l^{\prime}m^{\prime}j^{\prime} |\text{C}|lmj \rangle$ has different forms. For pulse scheme (1) i.e. $\Omega_c(x,t)=\Omega_c(t)\sin(x+\phi(t))$:
\begin{align}
&\langle l^{\prime}m^{\prime}j^{\prime} |\text{C}|lmj \rangle=\bigg\{\frac{1}{2}(\delta_{l^{\prime},l-1}+\delta_{l^{\prime},l+1})\otimes\langle m^{\prime}|\Omega_c(t)\sin{\phi(t)}|m\rangle\nonumber\\&+\frac{1}{2i}(\delta_{l^{\prime},l+1}-\delta_{l^{\prime},l-1})\otimes\langle m^{\prime}|\Omega_c(t)\cos{\phi(t)}|m\rangle\bigg\}\otimes(\delta_{j^{\prime}2}\delta_{j3}+\delta_{j^{\prime}3}\delta_{j2}).
\end{align}
For pulse scheme (2) i.e. ${\Omega_c(x,t)=\Omega_{c2}(t)e^{i(\phi_2(t)+x)}/2i -\Omega^0_ce^{-ix}/2i}:$
\begin{align}
&\langle l^{\prime}m^{\prime}j^{\prime} |\text{C}|lmj \rangle=\bigg\{\frac{1}{2i}\delta_{l^{\prime},l+1}\otimes\langle m^{\prime}|\Omega_{c2}(t)e^{i\phi_2(t)}|m\rangle-\frac{\Omega^0_c}{2i}\delta_{l^{\prime},l-1}\otimes \delta_{m^{\prime}m}\bigg\}\otimes\delta_{j^{\prime}3}\delta_{j2}+\nonumber\\
&\bigg\{-\frac{1}{2i}\delta_{l^{\prime},l-1}\otimes\langle m^{\prime}|\Omega_{c2}(t)e^{-i\phi_2(t)}|m\rangle+\frac{\Omega^0_c}{2i}\delta_{l^{\prime},l+1}\otimes \delta_{m^{\prime}m}\bigg\}\otimes\delta_{j^{\prime}2}\delta_{j3}.
\end{align} 
The spatio-temporal probability distribution of a Bloch-Floquet mode is,
\begin{align}
    |\psi|^2
  &=\sum_{l^{\prime}m^{\prime}lmj}c_{l^{\prime}m^{\prime}j}c_{lmj}e^{i (l-l^{\prime}) x+i (m-m^{\prime}) t},
\end{align}
where
$
   |\psi\rangle=\langle x ,t |u_{q,E_q}(x ,t )\rangle\rangle
$
with the fractional probability of being in spin $|i\rangle$ at $x$ and $t$ ($x\in[-\pi,\pi]$, $t\in[-\pi,\pi]$) given by,
\begin{align}
   \frac{|\langle j|\psi\rangle|^2}{|\psi|^2}
  =\frac{\sum_{l^{\prime}m^{\prime}lm}c_{l^{\prime}m^{\prime}j}c_{lmj}e^{i (l-l^{\prime}) x+i (m-m^{\prime}) t}}{\sum_{l^{\prime}m^{\prime}lmj}c_{l^{\prime}m^{\prime}j}c_{lmj}e^{i (l-l^{\prime}) x+i (m-m^{\prime}) t}}.
\end{align}

It is important~\cite{acki2016} to appropriately choose the number of plane waves $L$ and $M$ to be large enough to accurately represent the couplings between the dark-state channel and bright-state channels. We solve for the lowest few dozen eigenstates near zero energy of these sparse matrices with dimensions ${3(2L+1)(2M+1)\times3(2L+1)(2M+1)\sim10^5\times10^5}$ using the Arnoldi algorithm. We find that the solution converges with $M$ as low as 25, however for all our calculations we use $M\simeq210$.

\subsection{Bandstructure of $\hat{H}_{DS}(x,t)$}  
\label{trottersection}
In this subsection, we outline the method used to numerically solve for the Bloch-Floquet bandstructure of the dark-state channel ignoring non-adiabatic couplings to the bright-state channels:
\begin{align}
    \hat{H}_{DS}(x,t)=&\frac{\hat{p}^2}{2m}
	+W_{DS}(x,t)
\label{HeffBO}
\end{align}
where $\hat{H}_{{DS}}(x+\lambda/2,t)=\hat{H}_{{DS}}(x,t)$ and  $\hat{H}_{DS}(x,t+T)=\hat{H}_{{DS}}(x,t)$. Due to the nonlinear nature of $W_{DS}(x,t)$ solving for the bandstructure in the extended Hilbert space approach requires a 2D Fourier transform of $W_{DS}(x,t)$. Instead we solve for the bandstructure using the approach outlined in Refs.~\cite{Holthaus2015,Hanggi,Eckardt2017,Eckardt2015} where we evaluate the  time evolution operator over one Floquet period, $\hat{U}(T,0)$, and then diagonalize it. 

Making the Bloch ansatz, $|\psi_{{DS}}(x,t)\rangle=e^{iqx}|u_{q,{DS}}(x,t)\rangle$, the time-dependent Schrodinger equation in dimensionless units is
\begin{align}
\frac{\partial}{\partial t}|u_{q,{DS}}(x,t)\rangle=-\frac{i}{\omega_T}\overbrace{\bigg(\bigg(-i\partial_{x}+ q\bigg)^2
	+W_{DS}(x,t)\bigg)}^{\hat{H}_{{q,DS}}}|u_{q,{DS}}(x,t)\rangle.
\label{final}
\end{align}
We determine the time evolution operator for one Floquet period $\hat{U}_{q,{DS}}(2\pi+t_0,t_0)$~\cite{Eckardt2015,Eckardt2017,Holthaus2015} and equate that to the time evolution operator of an effective Floquet Hamiltonian $ e^{-\hat{H}^F_{q,DS}[t_0]T/\hbar}$ where $\hat{H}^F_{q,DS}[t_0]$ is defined at a Floquet gauge $t_0$~\cite{Bukov2015}. 

The expression for $\hat{U}_{q,{DS}}(2\pi,0)$ for $t_0=0$ is derived as follows:
\begin{align}
|u_{q,{DS}}(x,2\pi)\rangle&=\hat{U}_{q,{DS}}(2\pi,0)|u_{q,{DS}}(x,0)\rangle\nonumber\\
&=\mathcal{T}(e^{-i \int_0^{2\pi} \hat{H}_{{q,DS}}(x,t) dt/\omega_T})|u_{q,{DS}}(x,0)\rangle\nonumber\\
&=\bigg(\prod^L_{l=0} e^{-i \hat{H}_{{q,DS}}(x,l\Delta t)\Delta t/\omega_T}\bigg)|u_{q,{DS}}(x,0)\rangle\nonumber\\
&=\bigg(\prod^L_{l=0}\hat{S}_{q,l}  e^{-i \hat{E}_q(l\Delta t)\Delta t/\omega_T} \hat{S}^{-1}_{q,l}\bigg)|u_{q,{DS}}(x,0)\rangle\nonumber\\
\implies&\hat{U}_{q,{DS}}(2\pi,0)=\bigg(\prod^L_{l=0}\hat{S}_{q,l}  e^{-i\hat{E}_q(l\Delta t)\Delta t/\omega_T} \hat{S}^{-1}_{q,l}\bigg)=e^{-\hat{H}^{F}_{q,DS}[0]T/\hbar}
\label{timeevolve}
\end{align}
where $\mathcal{T}$ is the time-ordering operator, $L\Delta t=2\pi$  and $L$ is an integer number of time-steps. $\hat{S}_{q,l}$ is chosen to diagonalize $\hat{H}_{{q,DS}}(x,l\Delta t)$ at time $l\Delta t$: $\hat{S}^{-1}_{q,l}\hat{H}_{{q,DS}}(x,l\Delta t)\hat{S}_{q,l}=\hat{E}_q(l\Delta t)$ where $\hat{S}^{-1}_{q,l}\hat{S}_{q,l}=I^{2L+1}$.
Finally, we diagonalize $\hat{U}_{q,{DS}}(2\pi,0)$ in Eq.~\ref{timeevolve} to evaluate the Floquet eigenvalues and the eigenvectors~\cite{Holthaus2015,Eckardt2015,Eckardt2017}:
\begin{align}
    \hat{U}_{q,{DS}}(2\pi,0)&=e^{-i\hat{H}^{F}_{q,DS}[0]T/\hbar}=\sum^{2L+1}_{j=1}e^{-iE^{j}_q T/\hbar}|u^{j}_{q,{DS}}(x,0)\rangle\langle u^{j}_{q,{DS}}(x,0)|.
\end{align}
The Floquet eigenvalues $E^{j}_q$ are time-independent. For $\lambda/(2N)$-spaced lattices the Rabi pulses are the same for each $T/N$ sub-Floquet period and symmetry arguments were used to speed up the creation of the one-period Floquet evolution operator $\hat{U}_{q,{DS}}(2\pi,0)$~\cite{Holthaus2015}.

\section{Effect of two-photon detuning}
\label{twophotonsection}
\begin{figure}
\centering
\includegraphics[height=3in]{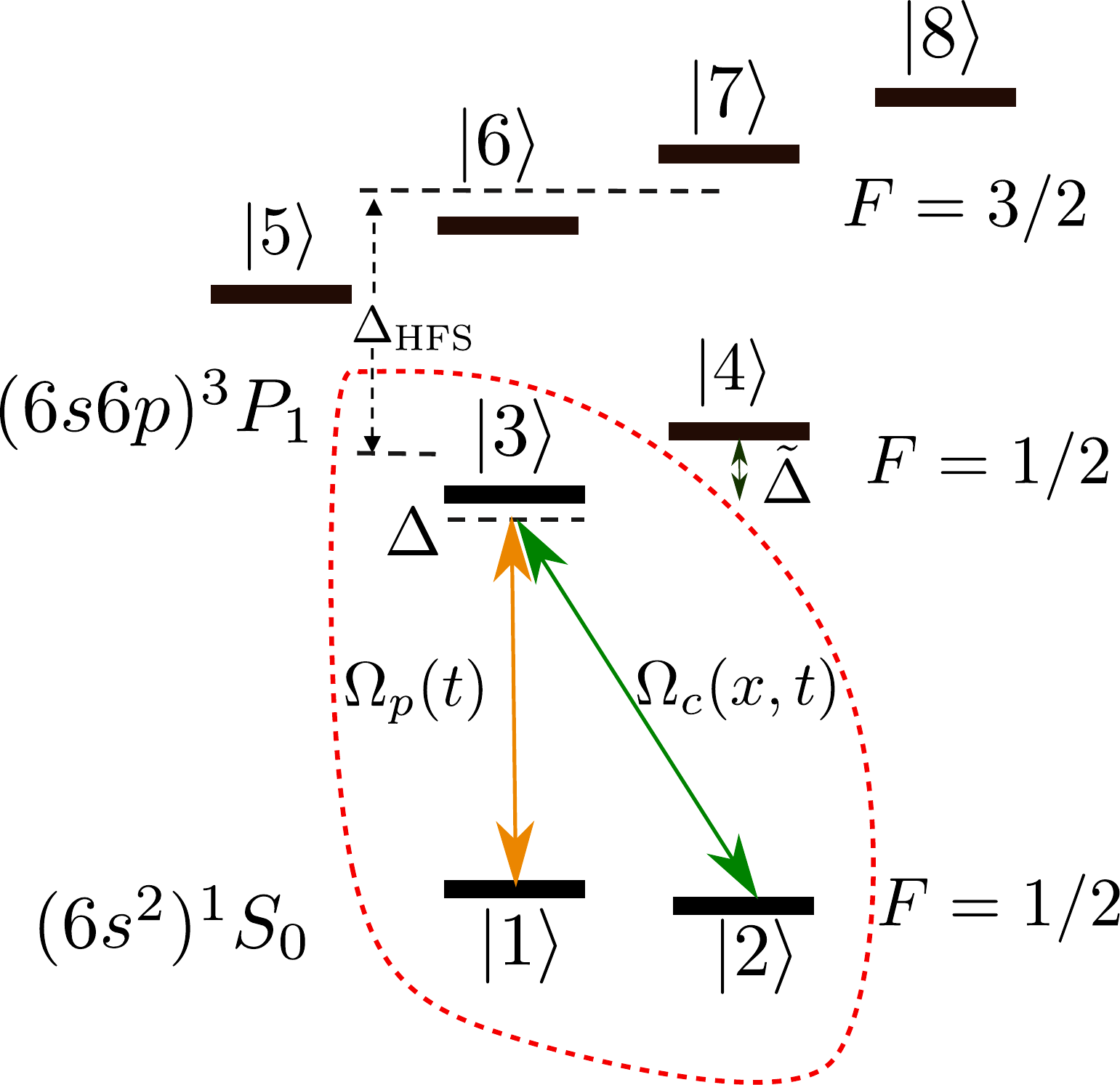}
 \caption{ Level structure of the $(6s^2)^1S_0$ and $(6s6p)^3P_1$ manifolds of $^{171}$Yb: $\Delta$ is the single photon detuning; $\tilde{\Delta}$ is the Zeeman splitting in the excited state due to an external magnetic field; and $\Delta_{\textrm{HFS}}$ is the $(6s6p)^3P_1$ hyperfine splitting. }
 \label{Ybscheme}
\end{figure}
In this section, we discuss the detrimental effect of the states outside the $\Lambda$-system in realizing $\lambda/(2N)$-spaced lattices for the specific case of $^{171}$Yb. In addition to the $\Lambda$-system composed of the states $|1\rangle$, $|2\rangle$ and $|3\rangle$, the ${(6s6p)}^3P_1$ manifold has 5 additional states   $|4\rangle$ to $|8\rangle$ that can couple to states $|1\rangle$ and  $|2\rangle$ (Fig.~\ref{Ybscheme}). When the effective Rabi frequencies are much smaller than their respective detunings to the off-resonant excited states $|4\rangle$ to $|8\rangle$, we can adiabatically eliminate the excited states and quantify their effect on the ground states $|1\rangle$ and $|2\rangle$ of the $\Lambda$-system in dimensionless units as follows:

\begin{align}
    \hat{\Omega}_{\text{OR}}(x,t)&\simeq\left(
\delta_1({t)}+\frac{\Omega_p^2(t)}{2 (\Delta_{\textrm{HFS}}+i\Gamma/2)}+\frac{3|\Omega_c(x,t)|^2}{8(\Delta_{\textrm{HFS}}+i\Gamma/2)}\right)|1\rangle\langle1|+\nonumber\\&\left(\delta_2({t)}+\frac{\Omega_p^2(t)}{4 (\tilde{\Delta}+i\Gamma/2)}+\frac{\Omega_p^2(t)}{2(\Delta_{\textrm{HFS}}+i\Gamma/2)}+\frac{|\Omega_c(x,t)|^2}{8(\Delta_{\textrm{HFS}}+i\Gamma/2)}\right)|2\rangle\langle2|\\
&\simeq\left(
\delta_1({t)}+\frac{\Omega_p^2(t)}{2 \Delta_{\textrm{HFS}}}+\frac{3|\Omega_c(x,t)|^2}{8\Delta_{\textrm{HFS}}}-i\frac{\Gamma_{1}(x,t)}{2}\right)|1\rangle\langle1|+\nonumber\\&\left(\delta_2({t)}+\frac{\Omega_p^2(t)}{4 \tilde{\Delta}}+\frac{\Omega_p^2(t)}{2\Delta_{\textrm{HFS}}}+\frac{|\Omega_c(x,t)|^2}{8\Delta_{\textrm{HFS}}}-i\frac{\Gamma_{2}(x,t)}{2}\right)|2\rangle\langle2|\label{twophotongen},\\&\textrm{where }\nonumber\\
 &\Gamma_{1}(x,t)=\Gamma\left( \frac{\Omega_p^2(t)}{2 \Delta^2_{\textrm{HFS}}}+\frac{3|\Omega_c(x,t)|^2}{8\Delta^2_{\textrm{HFS}}}\right),\label{nonherm1} \\
 &\Gamma_{2}(x,t)=\Gamma\left( \frac{\Omega_p^2(t)}{4 \tilde{\Delta}^2}+\frac{\Omega_p^2(t)}{2\Delta^2_{\textrm{HFS}}}+\frac{|\Omega_c(x,t)|^2}{8\Delta^2_{\textrm{HFS}}}\right), \label{nonherm2}
\end{align}
when $\Gamma\ll\tilde{\Delta},\Delta_{\textrm{HFS}}$.
By dynamically modulating $\delta_1(t)$ and $\delta_2(t)$ we  compensate for the spatially homogeneous but temporally modulated real parts in Eq.~\ref{twophotongen}. The compensated $\hat{\Omega}_{\text{OR}}(x,t)$ is added to $\hat{H}(x,t)$ and solved for using the method outlined in Sec.~\ref{blochfloquetbandstructuremain} and~\ref{genhamiltoniansection} to calculate the Bloch-Floquet bandstructure of $\lambda/(2N)$-spaced lattices in the presence of two-photon detunings and photon scattering loss due to states outside the $\Lambda$-system. The non-Hermitian terms in Eq.~\ref{twophotongen} (Eqs.~\ref{nonherm1} and~\ref{nonherm2}) account for loss from photon scattering due to admixing of the adiabatically eliminated excited states with the bare stable ground states $|1\rangle$ and $|2\rangle$. For a $\lambda/4$-spaced lattice created by pulse scheme (1), Eq.~\ref{twophotongen} is
		\begin{align}
&\langle l^{\prime}m^{\prime}j^{\prime} |\hat{\Omega}_{\text{OR}}(x,t)|lmj \rangle=-i\frac{\Gamma}{4\Delta^2_{\textrm{HFS}}}\delta_{l^{\prime},l}\nonumber\otimes\langle m^{\prime}|\Omega^2_p(t)|m\rangle\otimes\bigg(\delta_{j^{\prime}1}\delta_{j1}+\delta_{j^{\prime}2}\delta_{j2}\bigg)\\&-i\frac{\Gamma}{8\tilde{\Delta}^2}\delta_{l^{\prime},l}\nonumber\otimes\langle m^{\prime}|\Omega^2_p(t)|m\rangle\otimes\delta_{j^{\prime}2}\delta_{j2}-i\frac{\Gamma}{2\Delta^2_{\textrm{HFS}}}\delta_{l^{\prime},l}\nonumber\otimes\langle m^{\prime}|\Omega^2_c(t)|m\rangle\otimes\bigg(\frac{3}{16}\delta_{j^{\prime}1}\delta_{j1}+\frac{1}{16}\delta_{j^{\prime}2}\delta_{j2}\bigg)\\&-\left(\frac{1}{\Delta_{\textrm{HFS}}}-i\frac{\Gamma}{2\Delta^2_{\textrm{HFS}}}\right)\bigg\{\frac{1}{2}(\delta_{l^{\prime},l-2}+\delta_{l^{\prime},l+2})\nonumber\otimes\langle m^{\prime}|\Omega^2_c(t)\cos{2\phi(t)}|m\rangle\bigg\}\otimes\bigg(\frac{3}{16}\delta_{j^{\prime}1}\delta_{j1}+\frac{1}{16}\delta_{j^{\prime}2}\delta_{j2}\bigg),\\
\label{twophotongensimple}
\end{align}
where $\Delta_{\textrm{HFS}}\simeq-33,000~\Gamma$ and $\tilde{\Delta}\simeq-5500~\Gamma$ (Fig.~\ref{Ybscheme}). The real spatio-temporally modulated terms in Eq.~\ref{twophotongensimple} originate from $\Omega_c(x,t)$ and may only be compensated with the experimental overhead of adding more laser fields. 
\begin{figure}
\centering
\includegraphics[height=2in]{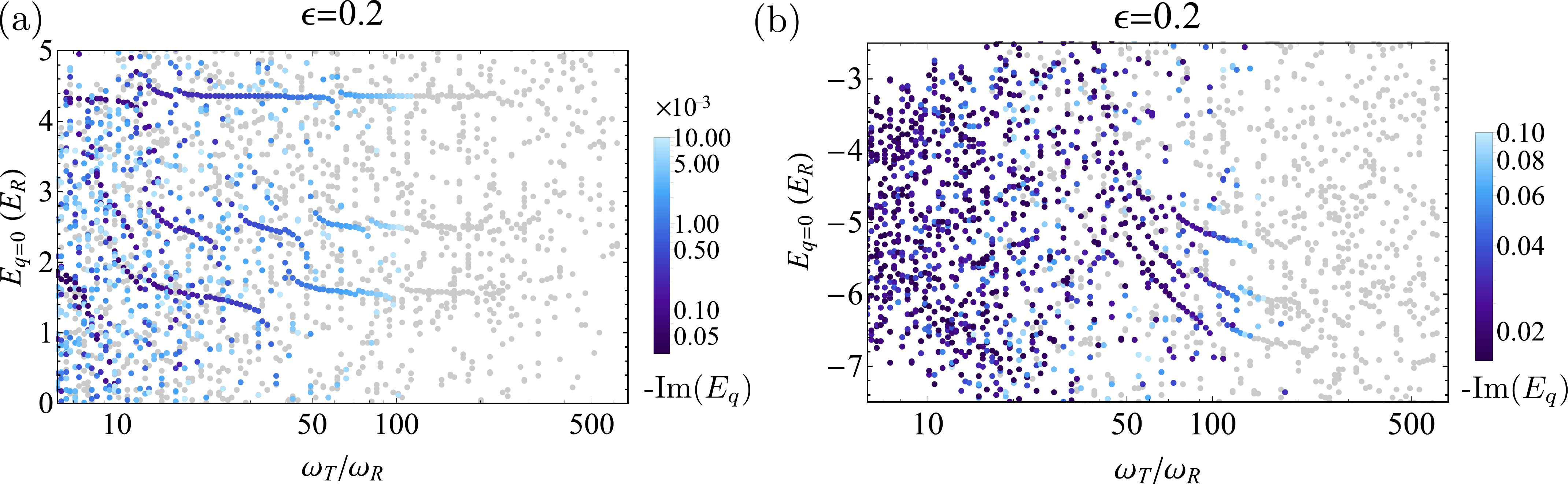}
 \caption{ The Floquet spectrum  $E_q$ at $q=0$ as a function of $\omega_T/\omega_R$ for a $\lambda/4$-spaced lattice: All calculations are performed for $\Omega^0_{ci}=600\Gamma$, $r_{0i}=0.01$, $\epsilon_i=0.2$, $\Delta=0$, and $t_{Si}=0.36T_{i}$. \textbf{(a)} For the ideal $\Lambda$-system \textbf{(b)} For the $\Lambda$-system with the spatio-temporally modulated ac-Stark shifts and losses due to the excited states $|4\rangle$-$|8\rangle$.}
   \label{contrasttwophoton}
\end{figure}

In Fig.~\ref{contrasttwophoton}{\color{magenta}b} we show the effect of $\hat{\Omega}_{\text{OR}}(x,t)$ and contrast it with an ideal $\Lambda$-system (Fig.~\ref{contrasttwophoton}{\color{magenta}a}). In Fig.~\ref{contrasttwophoton}{\color{magenta}b}, the window of operational $\omega_T$ is smaller and the losses are higher as expected. At lower $\omega_T$ the sinusoidal two-photon detunings are not time averaged out. The ground band is also red shifted due to the ac-Stark shifts being red detuned. Large Rabi frequencies with uncompensated spatio-temporally modulated two-photon detuning destroy the fidelity of the STIRAP pulses and make it harder to create $\lambda/(2N)$-spaced lattices. Losses from excited states admixing further shorten lifetimes.

Another possible candidate system that uses fine-structure states instead of hyperfine states for the $\Lambda$-system consists of the metastable states $(6s6p)^3P_2$ and $(6s6p)^3P_0$ of Yb as the long-lived states $|1\rangle$ and $|2\rangle$, and $(6s7s){}^3S_1$ as the excited state $|3\rangle$. This $\Lambda$-system is well isolated, and $|1\rangle$ and $|2\rangle$ are separated in energy by multiple THz. In this configuration, $\lambda/(2N)$-spaced lattices can be realized for both bosonic and fermionic species of Yb. The large matrix elements for the $(6s6p)^3P_2\leftrightarrow(6s7s){}^3S_1$ and $(6s6p)^3P_0\leftrightarrow(6s7s){}^3S_1$ transitions ensure that higher Rabi frequencies can be achieved in these systems without the detrimental effect of states outside the $\Lambda$-system. However Floquet heating from interactions~\cite{Eckardt2015,Eckardt2017} and losses from fine-structure collisions of $(6s6p)^3P_2$ atoms~\cite{Yamaguchi} could limit lifetimes in these systems.

\section{Adiabatic loading into the ground band}
\label{adiabaticloading}
There are a few ways to adiabatically load into the ground band of a $\lambda/4$-spaced lattice given that one has control over $\Omega_{c1}(t)$, $\Omega_{c2}(t)$, $\Omega_{p}(t)$, $\phi_{1}(t)$, and $\phi_{2}(t)$. We consider here a protocol in which the time-averaged potential is grown by periodically pulsing $\Omega_c(t)$ with a slowly varying envelope $f(t)$ with a timescale much slower than the motional degree of freedom: $\tilde{\Omega}_c(t)=f(t)\Omega_c(t)$. The pulse profile for $\Omega_c(t)$ is determined by Eq.~\ref{pulseshape2} for a given final $\epsilon$. For pulse scheme (1), $\tilde{\Omega}_p(t)$ along the ramp is determined by $\tilde{\Omega}_{p}(t)=\sqrt{\beta^2-\tilde{\Omega}^2_{c}(t)}$. The large and constant energy gap $\hbar\beta/2$ minimizes admixing of the dark-state channel with the bright-state channels. Under these conditions the loading is spin adiabatic because
\begin{equation}
\left|\frac{\partial \tilde{\alpha}(x_{h},t)}{\partial t}\right|<|\beta r(t)|<\beta r_0,
\end{equation}
for $0<f(t)<1$ and $\dot{f}(t)\ll1/T$, where $\tilde{\alpha}=\tan^{-1}[\tilde{\Omega}_p(t)/\tilde{\Omega}_c(x,t)]$. 
For pulse scheme (2), we propose the protocol $\tilde{\Omega}_p(t)=(1-f(t))\Omega^0_p$ and the pulse profile for $\Omega_{c2}(t)$ is one that creates the desired $\lambda/4$-spaced lattice for a chosen $\epsilon$ according to Eq.~\ref{onebeamdrive}.  $\tilde{\Omega}_p(t)$ is reduced down from an initial large value to its final value of $\Omega^0_p$.
This ensures that the energy gap $\hbar\beta/2$ at the nodes is lower at the end of the ramp than at the start, minimizing admixing of the dark-state channel with the bright-state channels along the ramp.

\bibliographystyle{unsrt}
\section*{References}
\bibliography{Floquettheorypaper}

\end{document}